\documentclass[aps,twocolumn,prx,longbibliography,floatfix,superscriptaddress,footnoteinbib]{revtex4-1}
\usepackage[english]{babel}

\usepackage{graphicx}
\usepackage{float}

\usepackage{amssymb,amsfonts,amsmath}
\usepackage{bm}

\usepackage{times}

\usepackage[colorlinks,bookmarks=false,citecolor=blue,linkcolor=blue,urlcolor=blue]{hyperref}
\usepackage{color}
\usepackage[all]{hypcap} 

\newcommand{\ep}{\varepsilon}

\def\bem{\begin{multline}}
\def\eem{\end{multline}}
\newcommand{\be}{\begin{equation}}
\newcommand{\ee}{\end{equation}}
\newcommand{\ben}{\begin{equation*}}
\newcommand{\een}{\end{equation*}}
\def\ba{\begin{aligned}}
\def\ea{\end{aligned}}
\newcommand{\bea}{\begin{eqnarray}}
\newcommand{\eea}{\end{eqnarray}}
\def\bes{\begin{subequations}}
\def\ees{\end{subequations}}
\def\bal{\begin{align}}
\def\eal{\end{align}}

\newcommand{\la}{\left\langle}
\newcommand{\ra}{\right\rangle}
\newcommand{\lv}{\left|}
\newcommand{\rv}{\right|}
\newcommand{\lb}{\left[}
\newcommand{\rb}{\right]}
\newcommand{\lp}{\left(}
\newcommand{\rp}{\right)}

\newcommand{\diag}{{\rm \, diag\,}}

\begin{document}

\title{Robustness of delocalization to the inclusion of soft constraints in long-range random models}

\author{P.~A.~Nosov}
\email[]{p.nosov1995@gmail.com}
 \address{Department of Physics, St. Petersburg State University, St. Petersburg 198504, Russia}
 \address{NRC Kurchatov Institute, Petersburg Nuclear Physics Institute, Gatchina 188300, Russia}
 \address{Max-Planck-Institut f\"ur Physik komplexer Systeme, N\"othnitzer Stra{\ss}e~38, 01187-Dresden, Germany }

\author{I.~M.~Khaymovich}
\email[]{ivan.khaymovich@pks.mpg.de}
 \address{Max-Planck-Institut f\"ur Physik komplexer Systeme, N\"othnitzer Stra{\ss}e~38, 01187-Dresden, Germany }

\begin{abstract}
Motivated by the constrained many-body dynamics, the stability of the localization-delocalization properties
to the inclusion of the soft constraints is addressed in random matrix models.
These constraints are modeled by correlations in long-ranged hopping with Pearson {correlation} coefficient different from zero or unity.
Counterintuitive robustness of delocalized phases, both ergodic and (multi)fractal, in these models is numerically observed and confirmed by the analytical calculations.
First, matrix inversion trick is used to uncover the origin of such robustness.
Next, {to characterize delocalized phases}
a method of eigenstate calculation, sensitive to correlations in long-ranged hopping terms, is developed for random matrix models and approved by numerical calculations and previous analytical ansatz.
The effect of the robustness of states in the bulk of the spectrum the inclusion of to soft constraints is generally discussed for single-particle and many-body systems.
\end{abstract}
\date{\today}

\maketitle
\section{Introduction}
Absence of thermalization in interacting many-body quantum systems has attracted significant interest
and boosted numerous studies of different possibilities to violate eigenstate thermalization hypothesis (ETH) both in static and driven systems.
The first and most developed way to do this is to randomize system parameters by including disorder. {This phenomenon is called many body localization (MBL)}~\cite{Basko06,gornyi2005interacting}. Like in single-particle case of Anderson localization~\cite{Anderson1958} disorder induces destructive interference and provokes emergent local integrals of motion~\cite{serbyn2013local,huse2013phenomenology} blocking the excitation transport.

\begin{figure}[t!]
\centerline{
\includegraphics[width=0.9\columnwidth]{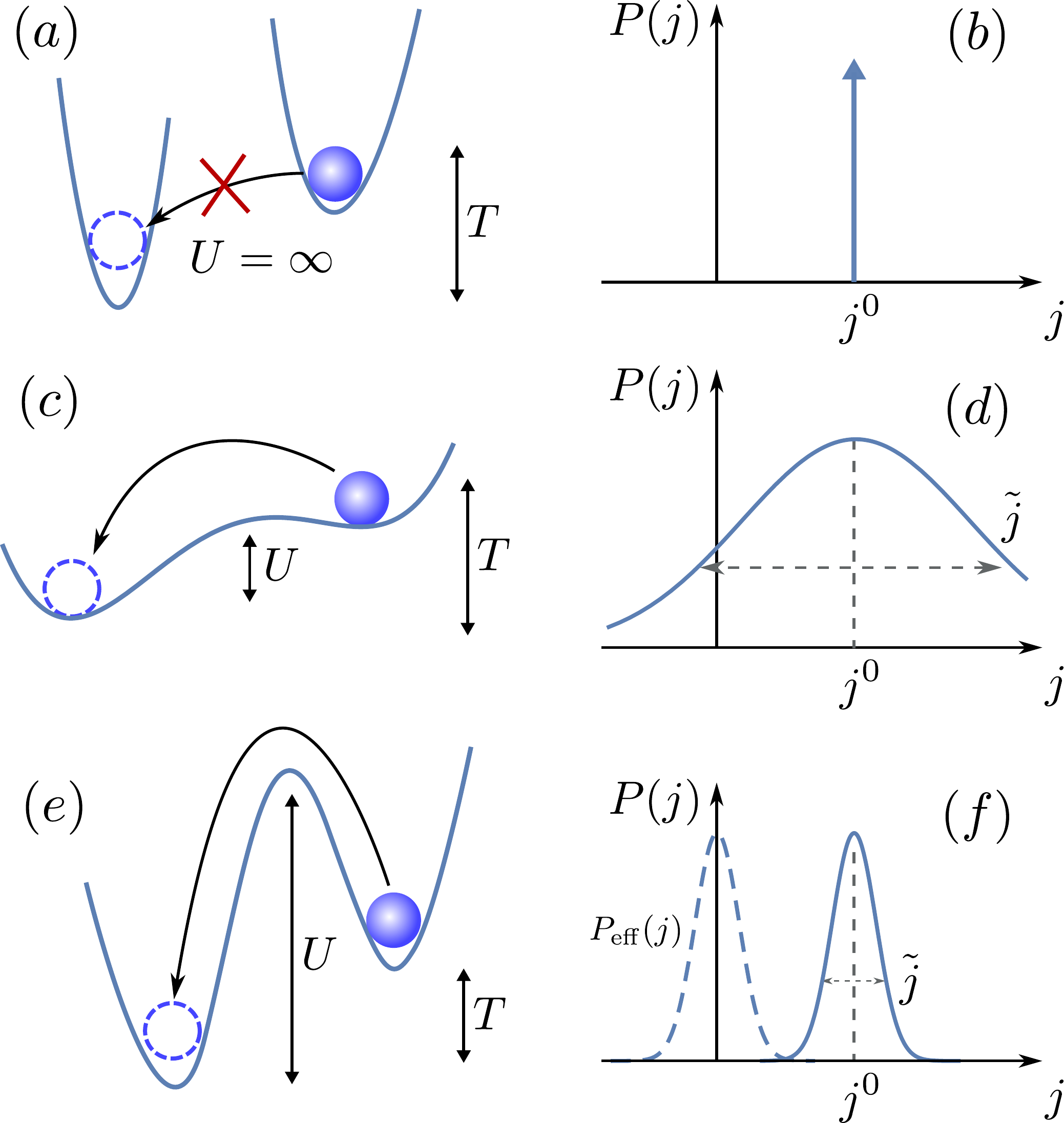}
}
\caption{{\bf Sketch of constraints}
in (a,~c,~e)~many-body systems and (b,~d,~f)~random matrix models.
Hard constraint corresponds to (a)~the infinite barrier $U=\infty$ between blocks of Hamiltonian (shown pictorially as parabolic potentials)
at any temperature~$T$
and (b)~the absence of any fluctuations $j^0/\tilde j=\infty$ in long-ranged hopping in random matrix analogs.
The opposite limit of very small constraint allows large fluctuations in both cases (c, d)
due to small ratios $U/T\ll 1$ and $j^0/\tilde j\ll 1$.
The most nontrivial case of soft constraint with finite but large barrier $U/T\gg 1$ and $j^0/\tilde j\gg 1$
(e)~provides the way to overcome the barrier with thermal-activated rate in many-body case, and
(f)~suggests that delocalization is determined by the width of the distribution ($P_{\rm eff}$), but not by its peak position ($P$).
}
\label{fig:P(j)}
\end{figure}

An alternative way to break ETH in absence of disorder is to add some hard constraints to the many-body system, that crucially reduce the Hilbert space by separating the Hamiltonian into the disjoint sub-block structure, see Fig.~\ref{fig:P(j)}(a). These hard constraints can be realized either by  infinitely strong interactions~\cite{Serbyn2018(1),Serbyn2018(2),Serbyn2018emergent,Khemani2019}, additional integrals of motion~\cite{Pollmann2019,Pretko2019(1),Pretko2019(2)}, or gauge invariance~\cite{Heyl2018gauge_fields}.
As a result, such hard constraints produce special low-entanglement states (such as many-body scars) in the bulk of the spectrum~\cite{Serbyn2018(1),Serbyn2018(2)}, giving significant contribution to the typical infinite-temperature states and revealing themselves via infinitely long-lived oscillations in quenched observables~\cite{Lukin2017}.
However, in real life none of barriers is infinite.
The effect of {\it soft} constraints on the thermalization in such systems
is non-trivial and under hot debate nowadays
as the finite-energy barriers between disjoint Hilbert space sub-blocks might be prevailed at high temperature, Fig.~\ref{fig:P(j)}(c,e).

For this reason it is of fundamental importance to study a simple model in which hard and soft constraints can be easily realized
and corresponding localization properties can be precisely investigated.
This would provide efficient criteria to characterize the effects of soft constraints in generic cases.

The straightforward analog of hard constraints in single-particle systems is given, e.g., by
fully-correlated long-ranged hopping in random matrix models~\cite{Burin1989,Yuzbashyan_JPhysA2009_Exact_solution,Yuzbashyan_NJP2016,Kravtsov_Shlyapnikov_PRL2018,Nosov2019correlation}.
Indeed, in these models the complete correlations of {\it all} hopping terms, Fig.~\ref{fig:P(j)}(b),
impose the hard constraints like in the case of many-body scars and localize the states in the bulk of the spectrum.
However, due to the single-particle nature of these systems, all states in the spectral bulk become localized~\cite{TI-models-footnote}.
Soft constraints in this case can be easily realized by considering partial correlations of long-ranged hopping with non-integer Pearson {correlation} coefficient. 
In this work we consider such single-particle disordered models with partially correlated long-ranged hopping, Fig.~\ref{fig:P(j)}(d,~f),
investigate both numerically and analytically their eigenstate statistics and phase diagrams,
and reveal unexpected robustness of delocalized phases to soft constraints, Fig.~\ref{fig:P(j)}(f).

In most random matrices the delocalized side of the localization transition (corresponding to ETH in many-body systems) is represented by Gaussian Wigner-Dyson ensembles~\cite{Mehta2004random}.
One of most well-known examples of a $d$-dimensional random matrix model confirming this statement and demonstrating the Anderson localization transition (ALT) at any $d$ (including $d\leq 2$) is
the power-law random banded matrix ensemble (PLRBM)~\cite{MirFyod1996},
\bes\label{eq:ham}
\begin{align}
\hat H = \hat \ep + \hat j , \quad
\hat \ep &= \sum_n \ep_n \left|n\ra\la n\right|, \\
\label{eq:ham_j}
\hat j&=\sum_{m,n}j_{mn} \left|m\ra\la n\right| \ ,
\end{align}
\ees
written in $d$-dimensional basis of $N$ lattice sites $\left|n\ra$.
This model is characterized by the independent Gaussian distributed hopping terms $j_{mn}$, with the standard deviation $\langle j_{mn}^{2}\rangle^{1/2}\propto |m-n|^{-a}$ power-law decaying at large distances $|m-n|\gg 1$ and
its ALT is governed by the decay exponent $a$ responsible for the ratio of the on-site disorder $\ep_n=H_{nn}$~\cite{On-site-disorder-footnote} and hopping terms.
The system shows ergodic (localized) wave-function statistics for $a<d$ ($a>d$), while
the ALT point $a=d$ 
is characterized by so-called nonergodic extended (multifractal) wave functions typical for the ALT phase diagram at the criticality~\cite{MF-PLRBM-footnote,PLRBM-Bogomolny-footnote}.

However recently there have been found several models showing the whole nonergodic extended phases, see, e.g.,~\cite{Kravtsov_NJP2015,Nosov2019correlation}.
The milestone random matrix example in this row is the Rosenzweig-Porter ensemble (RP)~\cite{RP}.
This nominally $1$d model has infinitely long-ranged independent Gaussian distributed hopping elements
with the $N$-dependent variance $\langle j_{mn}^{2}\rangle\propto N^{-\gamma}$ and
apart from the ALT transition at $\gamma=2$~\cite{Pandey,BrezHik,Guhr,AltlandShapiro,ShapiroKunz,Shukla2000,Shukla2005}, it exhibits
an ergodic transition (ET) at $\gamma=1$ from the ergodic phase ($\gamma<1$)
to a whole phase of nonergodic extended states ($1<\gamma<2$) characterized by a non-trivial fractal support set~\cite{KravtsovSupportSet,RRGAnnals} of wave functions~\cite{Kravtsov_NJP2015} with the fractal Hausdorff dimension $D=2-\gamma$, $0<D<1$.
This behavior has been further confirmed by several analytical and numerical papers~\cite{Biroli_RP,Ossipov_EPL2016_H+V,Amini2017,vonSoosten2017phase,vonSoosten2017non,Monthus2017,BogomolnyRP2018,RP_R(t)_2019,RP-like-footnote}.

The question of constraints (hopping correlations)
imposed in both above mentioned models has been considered recently in~\cite{Nosov2019correlation}.
Indeed, the new paradigm of the ALT suggested there
states that hopping correlations $\la j_{mn}j_{m'n'}\ra-\la j_{mn}\ra\la j_{m'n'}\ra\ne 0$
shrink in general an ergodic phase towards smaller disorder strengths extending both localized and multifractal phases.
In the case when all hopping integrals are fully-correlated (with unit Pearson's coefficient)
the localization at any disorder strength is restored~\cite{Burin1989,Kravtsov_Shlyapnikov_PRL2018,Borgonovi_2016} similar to the case of the short-range Anderson model in $d=1,2$~\cite{Anderson1958,abrahams1979scaling}.
An example of such random matrix models with fully-correlated hopping elements $j_{m\ne n} = C |m-n|^{-a}$, decaying
with the distance $|m-n|$ as a power-law like in PLRBM, has been suggested in a seminal paper by Burin and Maksimov (BM)~\cite{Burin1989}.
The infinitely long-ranged limit of this model (analogous to RP) with complete correlations between hopping terms $j_{m n} = C N^{-\gamma/2}$ 
has been shown to be exactly integrable by Yuzbashyan and Shastry (YS)~\cite{Yuzbashyan_JPhysA2009_Exact_solution,Yuzbashyan_NJP2016}.
Both these models demonstrate localization for all eigenstates, except measure zero, for all values of parameters $a$ and $\gamma$.
Note that the statistics of the site-independent scalar $C\sim 1$ does not play any role here.

A representative type of correlations considered in Ref.~\cite{{Nosov2019correlation}} is
the hard constraint (correlations with Pearson's coefficient $+1$) of {\it certain} pairs, $(m,n)$ and $(m',n')$, of hopping terms,
\be
j_{mn}/j_{m'n'} = f(m,n)/f(m',n')>0 \ .
\ee
Here $f(m,n)>0$ is the deterministic function of indices $m$, $n$, and possibly of the system size $N$.
For uncorrelated models (like PLRBM and RP) the pairs are only $(m',n') = (n,m)$, while in fully-correlated examples (BM and YS) all pairs of $(m',n')$ and $(m,n)$ are involved.
In the intermediate translation-invariant case $|m'-n'|=|m-n|$~\cite{TI-models-footnote}.

In this paper we address a complimentary aspect of soft constraints, namely
{\it partial} hopping correlations of {\it all} pairs of hopping terms
\be\label{eq:part_corr}
0<\la j_{mn}j_{m'n'}\ra<\sqrt{\la j_{mn}^2\ra\la j_{m'n'}^2\ra} \ .
\ee
For this we consider both PLRBM and RP models with {\it finite} hopping average values that interpolate between original uncorrelated PLRBM and RP ensembles and their fully-correlated counterparts, BM and YS models~\footnote{Here we take the scalar $C$ in fully-correlated models to be constant, as its statistics does not play any role,
and consider cross-correlations $< j_{mn}j_{m'n'}>$ instead of cross-covariances $< j_{mn}j_{m'n'}>-< j_{mn}><j_{m'n'}>\ne 0$ without loss of generality.}.

\begin{figure}[t!]
\centerline{
\includegraphics[width=1\columnwidth]{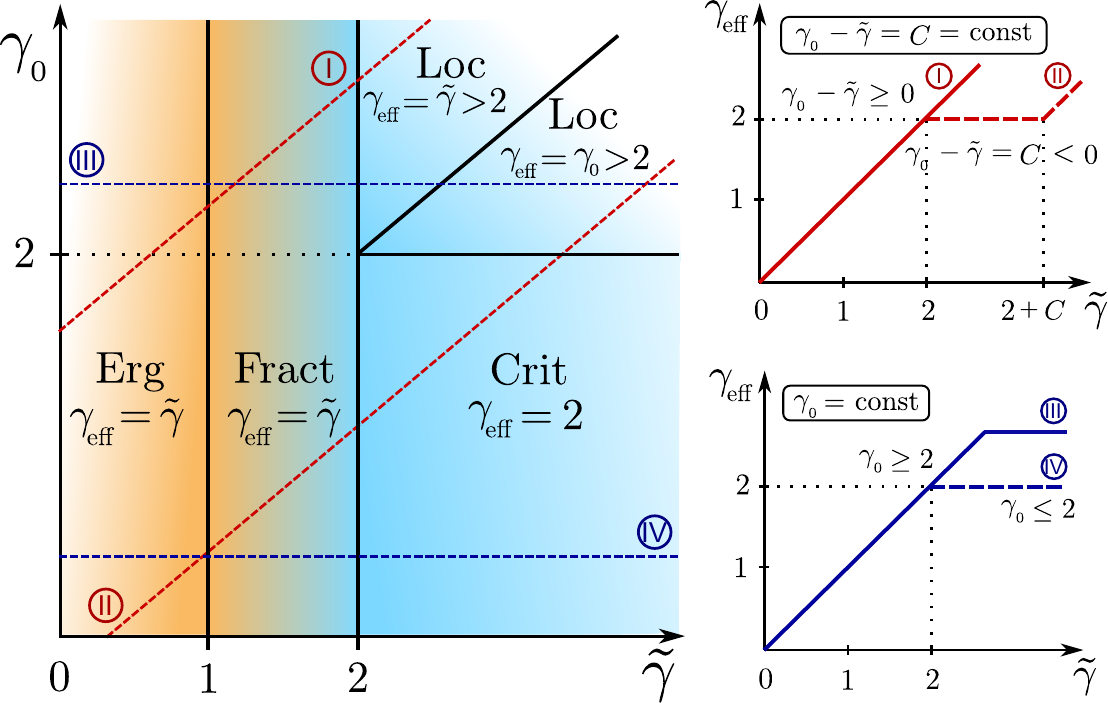}
}
\caption{{\bf Phase diagram of RP model with partial correlations~\eqref{eq:j_mn}}
with average hopping $j^0\sim N^{-\gamma_0/2}$ and its standard deviation $\tilde j \sim N^{-\tilde\gamma/2}$.
(left)~both the Anderson localization transition, $\gamma_{\rm eff}=2$, and the ergodic transition, $\gamma_{\rm eff}=1$,
are governed solely by $\tilde\gamma$, while $\gamma_{\rm YS} = \max(\gamma_0,2)$~\eqref{eq:gamma_eff_YS} affects wave-function profile only in the localized phase at $\gamma_{\rm YS}<\tilde \gamma$.
Right panels show the behavior of $\gamma_{\rm eff}$ along different cuts (I-IV) shown in left panel.
}
\label{Fig:gamma_phase_diagramm}
\end{figure}

The unexpected stability of the delocalized phases to soft constraints in both cases is demonstrated. 
The delocalization is shown to survive even for relatively narrow distributions with mean values $j^0$ much larger than the width $\tilde j$,
see Fig.~\ref{fig:P(j)}(f).
The positions of the ALT and possible ET are shown
to be governed solely by the distribution width $\tilde j$, but not by relative fluctuations $\tilde j / j^0$, see Fig.~\ref{Fig:gamma_phase_diagramm} and $P_{\rm eff}$ in Fig.~\ref{fig:P(j)}(f).
This brings us to the conclusion that soft constraints added to the initially delocalized phase
do not break delocalization of any state in the bulk of the spectrum, Fig.~\ref{fig:P(j)}(e),
even if the hard constraint does, Fig.~\ref{fig:P(j)}(a).

In order to uncover the origin of this counterintuitive result we first use the matrix inversion trick suggested in~\cite{Nosov2019correlation}
to rewrite the eigenproblem in the coordinate basis in an alternative way.
Furthermore we develop the self-consistent method of eigenvector calculation based on the averaging over off-diagonal matrix elements,
allowing one to access wave-function statistics and, in particular, confirming the phenomenological ansatz known in the literature for RP ensemble~\cite{Monthus2017,BogomolnyRP2018,RP_R(t)_2019} (see also~\cite{Ossipov_EPL2016_H+V}).
Unlike the standard renormalization group analysis~\cite{Levitov1990,MirFyod1996} or the Wigner-Weisskopf approximation~\cite{Monthus2017} used in the literature before this self-consistent method is sensitive to the hopping correlations.
In the current problem the full ALT diagram of previously mentioned models is calculated with help of these methods.

The rest of the paper is organized as follows.
In Sec.~\ref{Sec:Model} we formulate the random matrix models in the focus.
Sec.~\ref{Sec:numerics} shows how the naive guess for the behavior of these models fails,
and provide numerical results along with localization-delocalization phase diagrams.
Sec.~\ref{Sec:Matrix_inversion} describes the matrix inversion trick which
explains the behavior of these models and
allows us to uncover the origin of unexpected stability of delocalized phases.
In Sec.~\ref{Sec:Self-consist_method} we demonstrate the self-consistent method of eigenfunction calculation on the example of RP ensemble with finite mean hopping values.
In Conclusion we sum up or results and give an outlook.

\section{Models}\label{Sec:Model}
Throughout the text we focus on the generalized Anderson's single-particle model with long-ranged hopping terms,
represented by the Hamiltonian~\eqref{eq:ham}.
The uncorrelated diagonal disorder is given by independent identically distributed random on-site energies $\ep_n$ with zero mean and fixed variance
\be
\la \ep_n\ra=0 \ ,\quad \la\ep_n^2\ra = W^{2}.
\ee
The summation in \eqref{eq:ham_j} is taken over pairs of sites $m,~n$ coupled by
hopping integrals
\be\label{eq:j_mn}
j_{mn}=j_{nm}^* = j^0(|m-n|)+h_{mn} \tilde j(|m-n|) \ ,
\ee
characterized by the distant-dependent mean and standard deviation values, respectively
\be\label{eq:j_def}
\la j_{mn}\ra= j^0(|m-n|) \ ,\quad \sigma_j = \tilde j(|m-n|) \ ,
\ee
where $\sigma_j^2 = \la j_{mn}^2\ra-\la j_{mn}\ra^2$.
For simplicity we restrict our consideration to $d=1$, unless stated otherwise.
Here and further we denote by $h_{mn}$ i.~i.~d. random variables with zero mean $\la h_{mn} \ra = 0$ and unit variance $\la |h_{mn}|^2\ra=1$.

\begin{figure*}[t]
\centering
\includegraphics[width=0.75\textwidth]{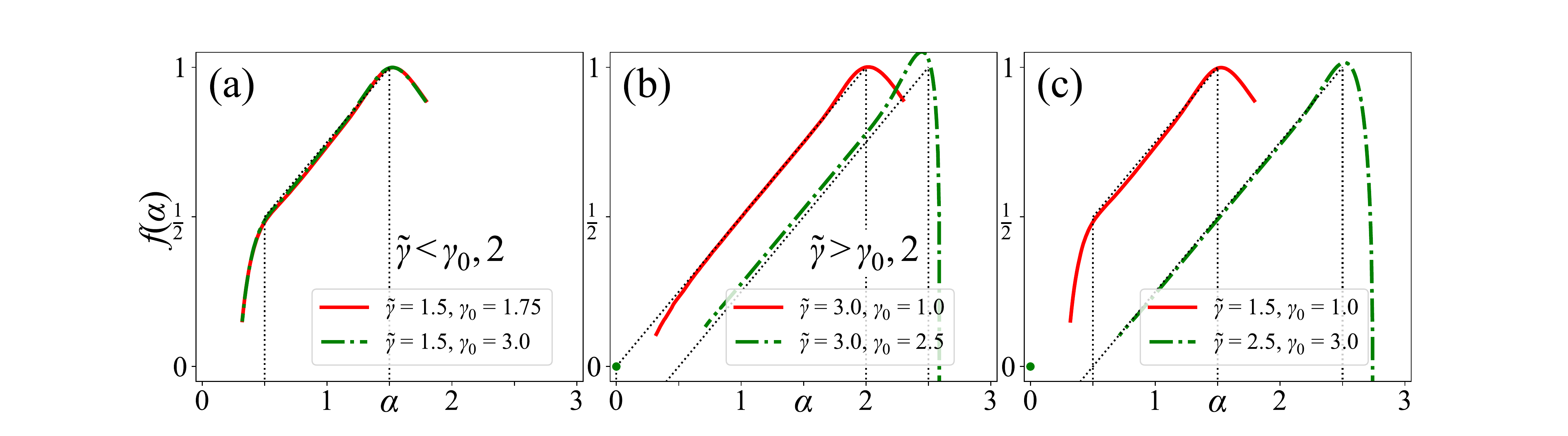}
\caption{{\bf Spectrum of fractal dimensions $f(\alpha)$ of RP model with partial correlations~\eqref{eq:j_mn}}
for different scaling of the average hopping $j^0\sim N^{-\gamma_0/2}$ and its standard deviation $\tilde j \sim N^{-\tilde\gamma/2}$. 
The finite-size data are numerically extrapolated to infinity from system sizes $N=2^{9}\ldots 2^{14}$ with $N_r = 10^3$ disorder realizations in each.
Dashed lines show analytical predictions~(\ref{eq:RP_f(a)}, \ref{eq:gamma_eff_RP+YS}) for $f(\alpha)$.
(a)~the naively expected case of small $\tilde \gamma<\gamma_0,2$ with fluctuations $\tilde j\gg j^0$ dominating over correlated hopping $j^0$
and determining the corresponding delocalized phase by $\gamma_{\rm eff} = \tilde \gamma$.
(b)~the case of large $\tilde \gamma>\gamma_0, 2$ where correlated hopping dominates $j^0\gg \tilde j$ and
localized the system with $\gamma_{\rm eff} = \gamma_{\rm YS} = \max(\gamma_0,2)\geq 2$.
(c)~non-trivial case $\min(\gamma_0,2)<\tilde \gamma<\max(\gamma_0,2)$
showing the dominant correlated hopping $j^0\gg \tilde j$, with the phase governed solely by $\gamma_{\rm eff} = \tilde \gamma$.
}
\label{Fig:f(a)_numerics}
\end{figure*}

\begin{figure*}[t]
\centering
\includegraphics[width=0.75\textwidth]{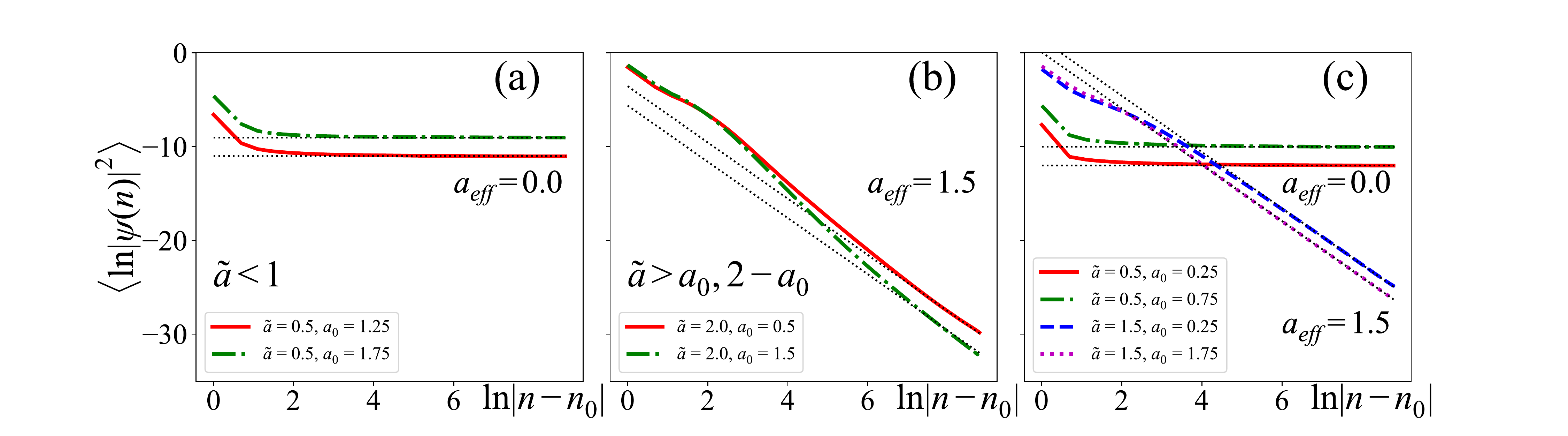}
\caption{{\bf Typical wave-function spatial decay
$\ln|\psi_{E}(n)|^{2}_{{\rm typ}} =\langle \ln(|\psi_E(n)|^{2}\rangle$ vs $n$
for PLRBM model with partial correlations~\eqref{eq:j_mn}} for different power-law decay rates of the average hopping $j^0_n\sim |n|^{-a_0}$ and its standard deviation $\tilde j_n\sim |n|^{-\tilde a}$. 
The data are numerically calculated for the system size $N=2^{14}$ and $N_r = 10^3$ disorder realizations.
Dashed lines show analytical predictions~(\ref{eq:PLRBM_psi_decay}, \ref{eq:a_eff_PLRBM+BM}) of this power-law decay.
Panels show the cases similar to ones in Fig.~\ref{Fig:f(a)_numerics}:
(a)~small $\tilde a<1,a_0$ with dominant fluctuations $\tilde j\gg j^0$  leading to the expected ergodic phase and $a_{\rm eff} = \tilde a$.
(b)~large $\tilde a>a_0, 2-a_0$ where correlated hopping dominates $j^0\gg \tilde j$ and
the localized phase is governed by $a_{\rm eff} = a_{\rm BM} = \max(a_0,2-a_0)$.
(c)~non-trivial case $\min(a_0,2-a_0)<\tilde a<\max(a_0,2-a_0)$
of dominant correlated hopping $j^0\gg \tilde j$ governed solely by $a_{\rm eff} = \tilde a$.
The data with $\tilde a<1$ in panels (a) and (c) is shifted with respect to each other for clarity.
}
\label{Fig:PLBM_WF_space_scaling}
\end{figure*}

The PLRBM and RP ensembles correspond to $j^0 = 0$ and
\be\label{eq:j_PLRBM+RP}
\tilde j_{PLRBM} = \frac{1}{|m-n|^{\tilde a}} \ \text{ and } \ \tilde j_{RP} = N^{-\tilde\gamma/2} \ ,
\ee
respectively, while for BM and YS models in contrast $\tilde j=0$,
\be\label{eq:j_BM+YS}
j^0_{\rm BM} = \frac{1}{|m-n|^{a_0}} \ \text{, and } \ j^0_{\rm YS} = N^{-\gamma_0/2} \ .
\ee

As infinitely long-ranged models (like RP and YS) do not have the notion of distance,
the main tool used to characterize the properties of their delocalized and localized phases
is the standard multifractal (MF) analysis.
This analysis is based on the spectrum of fractal dimensions~\cite{MirRev}
\be
f(\alpha) = 1-\alpha+\lim_{N\to\infty} \ln[P(|\psi_{E}(n)|^2=N^{-\alpha})]/\ln N \ ,
\ee
defined via the distribution of wave-function amplitudes $P(|\psi_{E}(n)|^2)$ with
the $N$-scaling of eigenfunctions $|\psi_{E}(n)|^2\propto N^{-\alpha}$, see Fig.~\ref{Fig:f(a)_numerics}.

Other long-ranged models (like PLRBM and BM) provide additional tools, e.g.,
the spatial decay of eigenfunctions with the distance $|n-n_0|$
from its maximal value at $n=n_0$~\cite{Kravtsov_Shlyapnikov_PRL2018,Nosov2019correlation}
given by the typical wave-function decay, see Fig.~\ref{Fig:PLBM_WF_space_scaling},
\be\label{eq:PLBM_WF_space_scaling}
|\psi_E(n)|^2_{typ}\equiv \exp\lb\la \ln|\psi_E(n)|^2\ra\rb \sim |n-n_0|^{-a_{\rm eff}} \ .
\ee
Here $\langle\ldots\rangle$ denotes the average over disorder and over eigenstates in the middle of the spectrum.
The energy level statistics (see, e.g.,~\cite{Bogomol2013}) as basis-invariant characteristics
gives the definite information about the fully-ergodic (Wigner-Dyson) phase and
the phase localized in a certain basis with the Poisson level statistics~\cite{Nosov2019correlation,r-stat-footnote}.

For RP-model the spectrum of fractal dimensions $f(\alpha)$ is shown to be linear in $\alpha = -\ln|\psi_{E}(n)|^2/\ln N$ for $\tilde\gamma\geq1$, with the slope $1/2$~\cite{Kravtsov_NJP2015}
\be\label{eq:RP_f(a)}
f(\alpha) = \left\{\begin{array}{lc}
1+(\alpha-\tilde\gamma)/2, & \alpha_{\min}<\alpha<\tilde\gamma \\
-\infty, & \alpha<\alpha_{\min} \text{ or }\alpha>\tilde\gamma\\
\end{array}
\right. \ ,
\ee
and an additional point $f(0)=0$ for $\tilde\gamma>2$.
Here $\alpha_{\min}=\max(0,\;2-\tilde\gamma)$.
The $f(\alpha)$ in the ergodic phase, $\tilde\gamma<1$, coincides with the one at $\tilde\gamma=1$ and is represented by the only  point $f(1)=1$~\cite{Kravtsov_NJP2015}.

The Yuzbashyan-Shastry (YS) ensemble (or as sometimes called the Type-1 model)~\cite{Yuzbashyan_JPhysA2009_Exact_solution,Yuzbashyan_NJP2016,Ossipov2013,Borgonovi_2016,Kravtsov_Shlyapnikov_PRL2018}
characterized by deterministic infinitely long-range hopping terms $j_{mn}=j^0 g_m g_n$, with the constants $g_m\propto O(1)$ of order of one,
is exactly integrable~\cite{Yuzbashyan_JPhysA2009_Exact_solution,Yuzbashyan_NJP2016}
and known to have all localized states for $j^0<1/N$ and all, except one, localized states for $j^0>1/N$ \cite{Yuzbashyan_NJP2016,Ossipov2013,Borgonovi_2016,Nosov2019correlation}.
The generalization of this ensemble to $N$-dependent hopping elements~\eqref{eq:j_BM+YS} shows a single-site localized phase for $\gamma_0>2$
and a critical behavior at $\gamma_0<2$
with the spectrum of fractal dimensions given by \eqref{eq:RP_f(a)} with $\gamma_0$ replaced by the following expression~\cite{Nosov2019correlation}
(see Appendix~\ref{App_Sec:YS} for analytical derivations)
\be\label{eq:gamma_eff_YS}
\gamma_{\rm YS} = \max(\gamma_0,\; 2)\geq 2 \ .
\ee


PLRBM undergoes the ALT at $a=1$, showing ergodic behavior at $a<1$ and power-law localization at $a>1$ \cite{MF-PLRBM-footnote,PLRBM-Bogomolny-footnote},
with the decay rate equal to the parameter $a$
\be\label{eq:PLRBM_psi_decay}
|\psi_{E}(n)| \propto |\psi_{E}(n_0)|/|n-n_0|^a
\ee
at the large distance $|n-n_0|\gg 1$ from the maximal point $n_0$,
$\max_n |\psi_{E}(n)| = |\psi_{E}(n_0)|$.

In BM-model~\cite{Burin1989,Malyshev2000,Malyshev_PRL2003,Malyshev2004,Malyshev2005,Borgonovi_2016,Kravtsov_Shlyapnikov_PRL2018,Nosov2019correlation}, the fully-correlated counterpart of PLRBM, which is determined by the Hamiltonian \eqref{eq:ham} with the hopping elements~\eqref{eq:j_BM+YS},
all, except measure zero 
of the states, are power-law localized \eqref{eq:PLRBM_psi_decay} in the entire region of the parameter $a$.
However, the power-law decay rate $a_{\rm BM}$ is not equal to $a$, but instead is always larger than one (see~\cite{Kravtsov_Shlyapnikov_PRL2018,Nosov2019correlation} for details)
\be\label{eq:a_eff_BM}
a_{\rm BM} = \max(a,\;2-a) \geq 1 \ .
\ee

\section{Intuitive guess and numerical results}\label{Sec:numerics}
What would be the phase diagram
for general models (\ref{eq:ham}~-~\ref{eq:j_def})
with both finite mean $j^0$ and fluctuating $\tilde j$ hopping terms?
For the first glance, it is natural to expect that
the behavior of uncorrelated models~\eqref{eq:j_PLRBM+RP} should be dominant as soon as $\tilde j\gg j^0$ ($\tilde \gamma<\gamma_0$ or $\tilde a<a_0$)
as the distribution of hopping elements is relatively wide and nearly centered at zero, see~Fig.~\ref{fig:P(j)}(d),
and vice versa the models with deterministic hopping~\eqref{eq:j_BM+YS} should dominate at $\tilde j\ll j^0$ ($\tilde \gamma>\gamma_0$ or $\tilde a>a_0$) when the distribution is relatively narrow and its width can be neglected, Fig.~\ref{fig:P(j)}(b,f).

However, this is not the case.
Indeed, from numerical calculations one can see that these models undergo the ALT (and the ET for RP case)
at the same points as their uncorrelated counterparts: $\tilde \gamma = 2$ ($\tilde \gamma=1$) and $\tilde a=1$
irrespective to the amplitude $j^0$ as if all mean values are zero, $j^0=0$, Fig.~\ref{Fig:gamma_phase_diagramm}.
Moreover, the wave-function statistics of such models in all phases, Figs.~\ref{Fig:f(a)_numerics} and~\ref{Fig:PLBM_WF_space_scaling},
coincides with the one of a simple mixture of two uncorrelated long-ranged hopping models
of the type~\eqref{eq:j_PLRBM+RP}, with hopping terms
\be\label{eq:j_tilde_mixture}
j_{mn} =  h_{mn}\tilde j(|m-n|) + h_{mn}^0\tilde j^0(|m-n|) \ ,
\ee
where the function $\tilde j^0(|m-n|)$ is described by \eqref{eq:j_BM+YS}
with the parameters $\gamma_0$ and $a_0$ of fully-correlated models (YS and BM)
replaced by their effective values \eqref{eq:gamma_eff_YS} and \eqref{eq:a_eff_BM}, respectively.
Thus, in the leading approximation one can replace the mixture of hopping elements by their maximum
\be
j_{mn} = h_{mn}\max\left[\tilde j(|m-n|), \tilde j^0(|m-n|)\right]
\ee
and map the model with partial correlations to the one with zero mean (equivalent to RP and PLRBM)
and parameters $\gamma$ and $a$ in Eqs.~(\ref{eq:RP_f(a)}, \ref{eq:PLRBM_psi_decay}) replaced by effective ones
\bes\label{eq:gamma_eff_a_eff}
\begin{align}\label{eq:gamma_eff_RP+YS}
\gamma_{\rm eff} &= \min(\tilde\gamma,\gamma_{\rm YS})= \min\left(\tilde\gamma,\max(\gamma_0,2)\right) \ , \\
\label{eq:a_eff_PLRBM+BM}
a_{\rm eff} &= \min(\tilde a, a_{\rm BM}) = \min\left(\tilde a, \max(a_0,2-a_0)\right)\ .
\end{align}
\ees
This is the main result of this paper shown here numerically, Figs.~\ref{Fig:f(a)_numerics} and~\ref{Fig:PLBM_WF_space_scaling}, and confirmed further analytically.

The qualitative explanation of this unexpected stability of delocalized states originates from the fact that
the models with deterministic long-range hopping terms (like YS and BM) demonstrate
only localized eigenfunction statistics in the bulk of the spectrum, but never delocalized.
As a result, in the mixture these fully-correlated models can compete with their uncorrelated counterparts
only in the localized phase, $\tilde a>1$ or $\tilde \gamma>2$, Fig.~\ref{Fig:gamma_phase_diagramm}, affecting the wave-function spatial profile
in the locator expansion~\cite{Anderson1958,Levitov1989,Levitov1990} (see also~\cite{Nosov2019correlation} for more detailed discussion).
In terms of soft constraints this means that
as soon as typical states in the spectral bulk are considered
the infinite temperature corresponding to them 
prevail over the finite barrier of soft constraint between previously disjoint sub-blocks of the Hamiltonian and
brings the system to the phase where it was before imposing constraints.

To understand the origin of the above mentioned behavior of partially-correlated models~\eqref{eq:part_corr},
summarized in Eqs.~(\ref{eq:j_tilde_mixture}~--\ref{eq:gamma_eff_RP+YS}),
in the next section we describe a matrix inversion trick~\cite{Nosov2019correlation}
providing an alternative representation of the eigenproblem and
apply it to the mixture of YS and RP model as an example.

\section{Matrix inversion trick}\label{Sec:Matrix_inversion}
Here we restrict our consideration of the matrix inversion trick to the case
of the mixture of RP and YS models (for the mixture of PLRBM and BM models please see Appendix~\ref{App_Sec:Matrix_inversion}).
For the first time this method has been suggested by us in Ref.~\cite{Nosov2019correlation}
to analytically prove the duality of the eigenfunction power-law decay in the $1$d BM-model ~\eqref{eq:a_eff_BM}
numerically discovered in Ref.~\cite{Kravtsov_Shlyapnikov_PRL2018} and to generalize both Anderson localization~\cite{Anderson1958,Levitov1989,Levitov1990} and Mott ergodicity~\cite{Mott1966} principles for the models with correlated hopping.
However recently there have been found several many-body~\cite{luitz2018emergent,Burin_Heyl} and higher-dimensional, $d>1$, single-particle models~\cite{Cantin2018,Deng_Burin},
applied to which this method easily uncovers their phase diagrams and the wave-function structure by the extension the locator expansion validity range.

\begin{figure}[t]
\centerline{
\includegraphics[width=0.8\columnwidth]{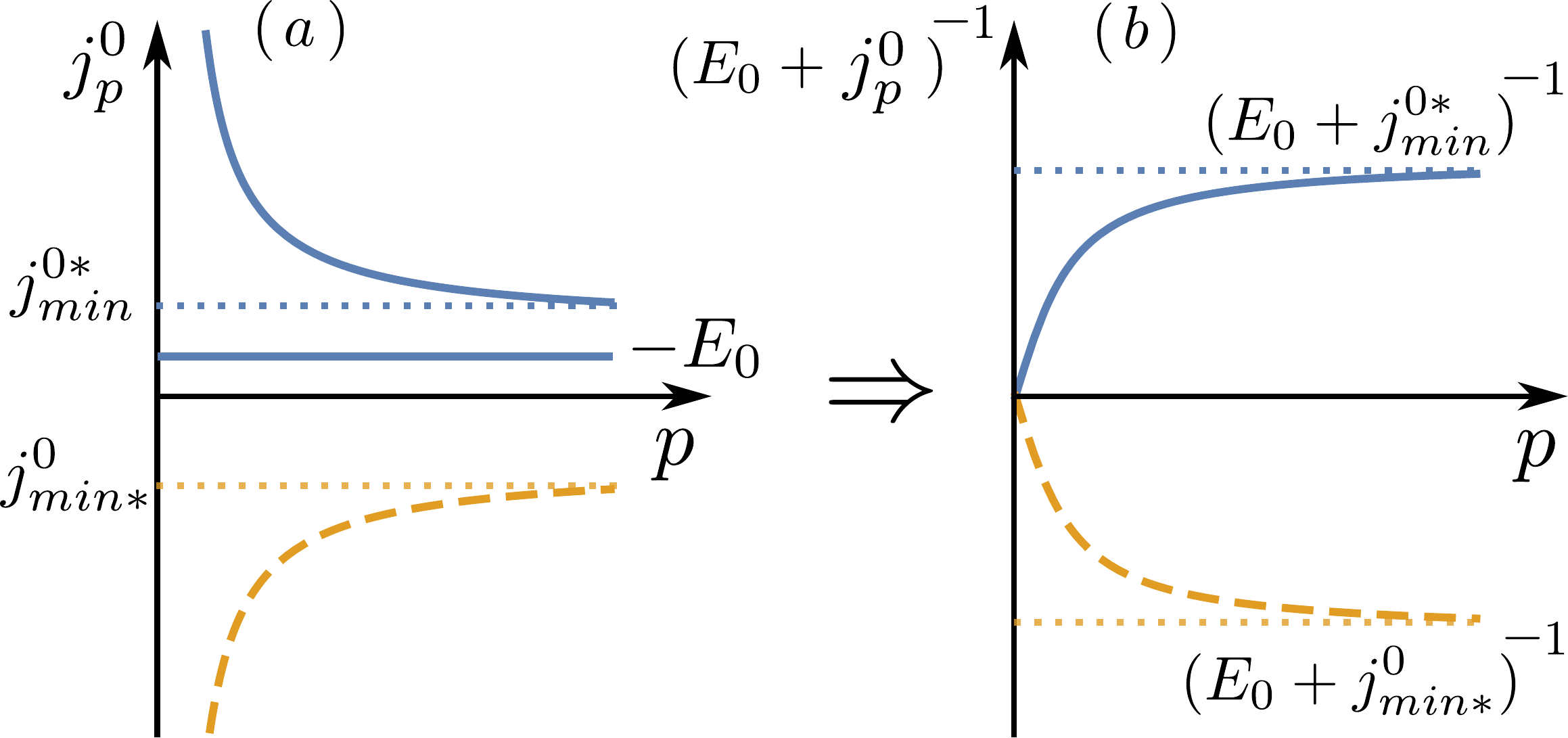}
}
\caption{{\bf Sketch of the spectrum of deterministic hopping~\eqref{eq:j_0},}
(a)~diverging from either (solid blue line) or both (blue and yellow lines) sides, but has a finite gap.
(b)~spectrum of the inverse matrix $\hat M$~\eqref{eq:M_matrix} with diminished divergence(s).
}
\label{Fig:j_p_spectrum}
\end{figure}

Let's first consider the pure deterministic (BM or YS) model~\eqref{eq:j_BM+YS}.
The matrix inversion trick is based on the spectral properties of the hopping matrix
\be\label{eq:j_0}
\hat j^0=\sum_{\la m,n\ra}j^0(|m-n|) \left|m\ra\la n\right|
\equiv \sum_p j^0_{p} \left|p\ra\la p\right| \
\ee
diagonalized in a certain basis $\left|p\ra$ (momentum basis for BM model).
If the spectrum $j^0_{p}$ of this matrix diverges from either side in the thermodynamic limit
(e.g., $j^0_{\max} = \max_p j^0_{p} \rightarrow \infty$ for $N\to \infty$ and $j^0_{\min} = \min_p j^0_{p}$ is finite)
or even from both sides, but has a finite spectral gap, see Fig.~\ref{Fig:j_p_spectrum}(a),
one can diminish the effect of these divergent terms to the hopping elements,
inverting the matrix $(1+\hat j^0/E_0)\equiv \hat M^{-1}$
\be\label{eq:M_matrix}
\hat M = (1+\hat j^0/E_0)^{-1} = \sum_p \frac{\left|p\ra\la p\right|}{1+j^0_{p}/E_0} \ .
\ee

This matrix inversion sends the diverging top-spectrum (or edge-spectrum) terms close to $j^0_{\max}$ to the denominator of the sum
while the condition $E_0+j^0_{\min}>0$ 
avoids the divergence of the contributions from the bottom (or states close to the gap) of the spectrum, see Fig.~\ref{Fig:j_p_spectrum}(b).
The optimization of $E_0\sim N^\beta$ over the parameter $\beta$~\cite{Nosov2019correlation}
gives the smallest effective hopping terms at $\beta=0$ in the whole parameter range (please see Appendix~\ref{App_Sec:Matrix_inversion} for details).

After the matrix inversion trick the problem takes the form
\be\label{eq:Matrix_inversion_0}
\left[(1+\hat j^0/E_0)^{-1}(E-\hat\ep+E_0)-E_0\right]\left|\psi_{E}\ra = 0 \ .
\ee
The diagonal part $M_0$ of the matrix $(1+\hat j^0/E_0)^{-1}_{m,n}=M_{m-n}$ forms effective on-site disorder $M_0 \ep_n$ and eigenvalue
$M_0 (E+E_0)-E_0$ of the problem, while the hopping terms are formed by $M_{m-n\ne0}(E-\ep_n + E_0)$.

The main idea behind this matrix inversion trick uses the fact that
the eigenstates with large hopping energies $|j^0_{p}|\gg 1$ are barely affected by the disorder $\hat \ep$.
Thus, they nearly coincide with those hopping matrix eigenstates $\left|p\ra$ that give the main contribution to \eqref{eq:j_0}.
All other eigenstates corresponding to small hopping energies $|j^0_{p}|\sim O(1)$ are orthogonal
to these large-energy states at the spectral edge and thus almost orthogonal to the main contribution to the hopping matrix given by them.
As a result the states in the bulk of the spectrum ``see'' the hopping terms $M_{m-n\ne0}(E-\ep_n + E_0)$
which are significantly reduced compared to the initial ones $j^0_{m-n}$.

For the case of YS model the rank-$1$ matrix $\hat j^0$ has the only non-zero eigenvalue $j^0_{p=0}=N^{1-\gamma_0/2}$
corresponding to the zero-momentum state $\la n| 0\ra_p=N^{-1/2}$
with arbitrary rest basis states $\left| g_{k\ne0}\ra$ orthonormal to $\left| 0\ra_p$~\footnote{For the general YS model the latter should be modified by $<n| g_0>=g_n/G$, with $G^2 = \sum_n |g_n|^2$.}.

The same matrix inversion trick \eqref{eq:Matrix_inversion_0} can be applied as well to the model~\eqref{eq:j_def} with partial correlations~\eqref{eq:part_corr}
\be\label{eq:Matrix_inversion}
\left[(1+\hat j^0/E_0)^{-1}(E-\hat\ep-\hat{\tilde{j}}+E_0)-E_0\right]\left|\psi_{E}\ra = 0 \ ,
\ee
where we inverse only the deterministic hopping part with the semi-infinite spectrum.


As a result Eq.~\eqref{eq:Matrix_inversion} takes the predicted form of Eq.~\eqref{eq:j_tilde_mixture}
\be\label{eq:Matrix_inversion_result}
\left[\hat\ep+\hat{\tilde{j}}+\hat{\tilde{j}}^0+\hat r\right]\left|\psi_{E}\ra = E\left|\psi_{E}\ra \ .
\ee
with the fluctuating elements of the matrix $\hat{\tilde{j}}^0$
\be\label{eq:tilde_j_0}
\tilde{j}^0_{mn} = -\frac{E-\ep_n+E_0}{N+E_0 N^{\gamma_0/2}}\sim N^{-\gamma_{\rm YS}/2} \ ,
\ee
which do not break the locator expansion
and the residual term $\hat r$ small compared to $\hat{\tilde{j}}$
\be
r_{mn} = \frac{\sum_l \tilde{j}_{ln}}{N+E_0 N^{\gamma_0/2}}\sim \frac{N^{(1-\tilde\gamma)/2}}{N+E_0 N^{\gamma_0/2}}\ll \tilde{j}_{mn} \ .
\ee
In Eq.~\eqref{eq:tilde_j_0} $\gamma_{\rm YS}$ is given by~\eqref{eq:gamma_eff_YS}.
This derivation confirms our numerical observation~\eqref{eq:j_tilde_mixture} and concludes this section.
The consideration of the mixture of PLRBM and BM models 
is addressed in Appendix~\ref{App_Sec:Matrix_inversion}.

To sum up, in this section we have shown that
the effective locator expansion result~\cite{Nosov2019correlation} analogous to~\eqref{eq:PLRBM_psi_decay} is applicable in the case of the RP-YS mixture to calculate the wave function
in the whole localized phase
coinciding with the one of the uncorrelated model ($\tilde \gamma>2$) at any $\gamma_0$, Fig.~\ref{Fig:gamma_phase_diagramm}.

To calculate the wave-function statistics in {\it all} phases, including delocalized ones, in the next section we develop a self-consistent method sensitive to hopping correlations.

\section{Self-consistent eigenstate calculation}\label{Sec:Self-consist_method}
In this section we consider the self-consistent method of the wave-function calculation,
generalizing the perturbation theory. For simplicity we restrict our consideration to the mixture of RP and YS ensembles.
For more general analysis please see Appendix~\ref{App_Sec:self-consist}.
Separating hopping terms $j_{mn} = j^0 + h_{mn}\tilde j$~\eqref{eq:j_mn} in the Hamiltonian~\eqref{eq:ham}
into deterministic $j^0=j^0_{YS}=N^{-\gamma_0/2}$~\eqref{eq:j_BM+YS} and fluctuating $\tilde j=\tilde j_{RP}=N^{-\tilde\gamma/2}$~\eqref{eq:j_PLRBM+RP} parts in the eigenproblem
\be
(E_n-\ep_{k}) \psi_{E_n}(k) = j^0\sum_{l}\psi_{E_n}(l)+\tilde j\sum_{l}h_{kl}\psi_{E_n}(l) \ ,
\ee
one can formally write its solution
\bes\label{eq:psi_n(m)_dE_n_solution}
\begin{align}
\psi_{E_n}(k) &= \psi_{E_n}(n)\frac{J_{kn}}{\ep_n - \ep_k + J_{nn}} \ , \\
\label{eq:dE_n}
J_{kn} &\equiv a_n+P_{kn}+\tilde j  h_{kn}
 \ .
\end{align}
\ees
in terms of the sums
\bes\label{eq:a_Ps_def}
\begin{align}
a_n &= j^0\sum_{l} \psi_{E_n}(l)/\psi_{E_n}(n) \ ,\\
P_{kn} &= \tilde j \sum_{l\ne n} h_{kl}\psi_{E_n}(l)/\psi_{E_n}(n)
\end{align}
\ees
and $E_n - \ep_n=J_{nn}$ given by \eqref{eq:dE_n}, that should be calculated self-consistently. Here and further we choose the index $n$ in the energy $E_n$ in such a way that in absence of off-diagonal elements $j^0 = \tilde j = 0$ the wave function is localized at $k=n$, $\psi_{E_n}(k) = \delta_{k,n}$, with the energy $E_n = \ep_n$.

Averaging \eqref{eq:psi_n(m)_dE_n_solution} over the hopping elements $h_{mn}$ with fixed bare energies $\ep_n$ gives the following expression
for the wave-function intensity
\be\label{eq:psi2_in_P_mn}
\la\left|\psi_{E_n}(k)\right|^2\ra_{h_{kn}} \sim  \frac{\la|\psi_{E_n}(n)J_{kn}|^2\ra}{(\ep_n-\ep_k+\Delta E_n)^2 + \Gamma_{n}^2} \ .
\ee
Here we assume that the sums $P_{mn}$ and $a_n$ are self-averaging and thus they are uncorrelated from each other and from $h_{kl}$.
For RP model itself this approximation of self-averaging over hopping terms has been used in several papers~\cite{Biroli_RP,Ossipov_EPL2016_H+V} and confirmed there by other methods.
The fluctuating energy shift $J_{nn} = E_n - \ep_n$ after this averaging leads both to the energy shift $\Delta E_n$ and the level broadening $\Gamma_n$ in Eq.~\eqref{eq:psi2_in_P_mn}.
Both $\Delta E_n$ and $\Gamma_n$ are of the same order as $E_n - \ep_n$ and in principle contain contributions from all cumulants of $P_{nn}$ and $a_n$,
however for our analysis it is enough to consider only mean values and the first-order perturbation theory term $\tilde j h_{nn}$
with standard deviation $\tilde j$ taken into account
(for further details see Appendix~\ref{App_Sec:self-consist})
\be\label{eq:Gamma_n}
\Gamma_n\simeq \la a_n\ra +\la P_{nn}\ra+\tilde j \ .
\ee
Here $\Gamma_n$ plays a role of the level broadening.
This level broadening determines the size of the miniband of almost fully-correlated eigenfunctions like in RP-model~\cite{Kravtsov_NJP2015,RP_R(t)_2019}.
The factor $\la|\psi_{E_n}(n)J_{kn}|^2\ra$ in the numerator of Eq.~\eqref{eq:psi2_in_P_mn} guarantees the wave-function normalization and not important for the wave-function statistics.

Focusing on the $N$-scaling of $\Gamma_n \sim \Delta E_n$ one can show that
the localized state realizes at $\Gamma_n$ smaller than the mean level spacing $\delta\simeq W/N$ of the model without hopping,
the ergodic state corresponds to $\Gamma_n$ large compared to the bare band of the system $W\sim O(1)$,
while the fractal phase appears at intermediate values:
\be\label{eq:Gamma_scaling}
\begin{array}{ll}
\;\phantom{\delta\ll}\Gamma_n\ll \delta        & \Leftrightarrow \text{ localized phase}\\
\delta\ll\Gamma_n\ll N\delta\sim W             & \Leftrightarrow \text{ fractal phase}\\
\;\phantom{\delta\ll}\Gamma_n\gg N\delta & \Leftrightarrow \text{ ergodic phase}\\
\end{array}
\ee

In order to estimate the scaling of the level broadening and identify the corresponding phases one can make use of the self-consistent equations for $P_{kn}$ and $a_n$ which could be obtained by substituting expressions \eqref{eq:psi_n(m)_dE_n_solution} for $\psi_{E_n}(k)$ and $E_n$
 to \eqref{eq:a_Ps_def}. The resulting equations read as
\bes\label{eq:a_Ps_self-consist}
\begin{align}
a_n &= j^0\lp1+\sum_{l\ne n} \frac{a_n + \tilde j h_{ln} + P_{ln}}{\ep_n - \ep_l + J_{nn}}\rp \ ,\\
P_{kn} &= \sum_{l\ne n} \frac{\tilde j h_{kl}(a_n + P_{ln} + \tilde j h_{ln})}{\ep_n - \ep_l + J_{nn}}
 \ .
\end{align}
\ees

The next essential step is to average these equations over $h_{nm}$ relying on the above-mentioned self-averaging properties of these sums. As a result, Eqs.~\eqref{eq:a_Ps_self-consist} take the form 
\bes\label{eq:a_Ps_mean}
\begin{align}
\la P_{kn} \ra &= \tilde j^2  S_1 \delta_{k,n}\ , \\
\la a_n\ra &= \frac{1}{(j^0)^{-1}-S_1} \sim \min\lp j^0, S_1^{-1}\rp\ ,
\end{align}
\ees
with the sum $S_1$ given by
\be\label{eq:S_1,2}
S_1 = \sum_{l} \frac{1}{\ep_n - \ep_l + \Gamma_n} \ .
\ee
The latter can be calculated, e.g., in two limiting cases of (i)~the completely rigid spectrum $\ep_n -\ep_m = (n-m)\delta$, with $\delta$ being a bare mean level spacing without hopping, and
(ii)~the Poisson statistics of $\ep_n$. Up to prefactors unimportant for the multifractal analysis $S_1$ takes the form
\be\label{eq:S_1_res}
S_1(\Gamma_n) \sim \left\{
\begin{array}{ll}
1/\delta & \;\phantom{\delta\ll}\Gamma_n\ll N\delta \\
N/\Gamma_n & \;\phantom{\delta\ll}\Gamma_n\gtrsim N\delta \\
\end{array}
\right. \ .
\ee

Considering cases \eqref{eq:Gamma_scaling} and substituting~\eqref{eq:S_1_res} into~\eqref{eq:a_Ps_mean} and~\eqref{eq:Gamma_n} one easily obtains the result up to prefactors $O(1)$
\be\label{eq:Gamma_n_res}
\Gamma_n \sim \left\{
\begin{array}{ll}
N^{-\min(\tilde\gamma,\gamma_{\rm YS})/2} & \;\phantom{1<}\tilde \gamma>2, \text{ localized} \\
N^{1-\tilde\gamma} & 1<\tilde \gamma<2, \text{ fractal} \\
N^{(1-\tilde\gamma)/2} & \;\phantom{1<}\tilde \gamma<1, \text{ ergodic} \\
\end{array}
\right. \ ,
\ee
with $\gamma_{\rm YS} = \max(\gamma_0,2)$ given by~\eqref{eq:gamma_eff_YS}.

Comparison of the resulting $\Gamma_n$ with the one of the RP model shows the same $N$-scaling with the parameter $\tilde\gamma$ replaced by
$\gamma_{\rm eff}$ from Eq.~\eqref{eq:gamma_eff_RP+YS}.
One can also check it by the direct calculation of the multifractal spectrum.
Indeed, the resulting approximate wave function \eqref{eq:psi2_in_P_mn} scales as
\be
\la\left|\psi_{E_n}(k)\right|^2\ra_{h_{kn}} \sim \frac{N^{-\tilde\gamma}+N^{-\gamma_{\rm YS}}}{N^{-p}+\Gamma_n^2} \ ,
\ee
where the scaling of $\Delta E_n \sim \Gamma_n$ is given by~\eqref{eq:Gamma_n_res}
and we parameterized $(\ep_n-\ep_k)\sim N^{-p/2}$ with the parameter $p$.
For a certain distribution of the diagonal terms $\ep_n$ with the width $W\sim O(1)$ the scaling of the probability reads as
\be
P(\ep_n-\ep_k)d\ep_n = P(N^{-p/2})N^{-p/2}dp \sim N^{-p/2}dp
\ee
for positive $p>0$ as $P(\ep_n\to 0) \sim O(1)$.
For negative $p<0$ the probability is at least exponentially small in $\ep_n$ and thus, for multifractal analysis one should neglect it
focusing on $p\geq 0$.
As a result the spectrum of fractal dimensions 
\be
N^{f(\alpha)-1}d\alpha = P\left(\left|\psi_{E_n}\right|^2\sim N^{-\alpha}\right)d(\left|\psi_{E_n}(k)\right|^2)
\ee
can be found from the expression
\begin{multline}
N^{f(\alpha)-1}d\alpha=dP\lp N^{-\alpha} \sim \frac{N^{-\tilde\gamma}+N^{-\gamma_{\rm YS}}}{N^{-p}+\Gamma_n^2}\rp \\
= \max_{p>0}N^{-p/2}d\alpha \ ,
\end{multline}
where the maximization in r.h.s. is taken with respect to the condition $N^{-\alpha} \sim (N^{-\tilde\gamma}+N^{-\gamma_{\rm YS}})/({N^{-p}+\Gamma_n^2})$.
The maximal probability is given by the condition $0<p<-2\ln \Gamma_n/\ln N$ leading to $p = \gamma_{\rm eff} - \alpha$
and to the result~\eqref{eq:RP_f(a)} with $\gamma_{\rm eff}$ given by~\eqref{eq:gamma_eff_RP+YS}.
Moreover the boundaries $0<p<-2\ln \Gamma_n/\ln N$ provide the correct bounds for $\alpha$
\be
\max(0,2-\gamma_{\rm eff})<\alpha<\gamma_{\rm eff} \ .
\ee
This analysis confirms numerical results shown in Fig.~\ref{Fig:f(a)_numerics} and concludes this section.
Note that the method developed in this section is powerful and accurate for the multifractal analysis
as one can take into account cumulants of any order of the sums~\eqref{eq:a_Ps_self-consist} fluctuating with the hopping terms
and sensitively distinguish models with slightly different hopping correlations.

\section{Conclusion and discussions}
To sum up, in this work we address
the effect of soft constraints on the phase diagram of
random matrix models with long-ranged correlated hopping.
We demonstrate unexpected robustness of the delocalized phases to partial hopping correlations imposed by soft constraints
and determine wave-function statistics and corresponding phase diagrams of milestone disordered long-range models,
power-law banded random matrix and Rosenzweig-Porter ensembles.
This main result~\eqref{eq:gamma_eff_a_eff} is confirmed by both numerical calculations and two analytical approaches.
The matrix-inversion trick developed in~\cite{Nosov2019correlation} uncovers the effective Hamiltonian~\eqref{eq:Matrix_inversion_result}
and confirms the wave-function behavior in the localized phase.
The self-consistent method allows to calculate wave-function statistics in delocalized phases as well and confirm the main result of the paper.

A parallel drown between constrained random-matrix models and many-body systems,
brings us to the conclusion that in general soft constraints added to the initially delocalized phase
do not break delocalization of any typical (infinite temperature) state,
even if the hard constraint does.
Indeed, the infinite temperature corresponding to the typical states in the spectral bulk
prevail over the finite barrier of soft constraint between previously-disjoint sub-blocks of the Hamiltonian and
brings the system to the phase where it was before imposing constraints.
However, the relations of slow-dynamics phenomena~\cite{lezama2019apparent}
to hard and soft constraints both in many-body systems~\cite{Heyl2018gauge_fields,Serbyn2018(1),Serbyn2018(2),Serbyn2018emergent,Khemani2019,Pollmann2019,Pretko2019(1),Pretko2019(2),torres2015dynamics,tavora2017power,torres2018generic,santos2018nonequilibrium}
and in closely related single-particle disordered models~\cite{RP_R(t)_2019, RRG_R(t)_2018} is still under debates and consideration.
The question of the dynamics and relaxation of highly non-local operators~\cite{khaymovich2019eigenstate} is also in the focus of the research in the community.

Another intriguing question is how the interplay between partial correlations in hopping and interaction amplitudes could affect localization properties in many-body systems with either or both hopping and interaction terms being long-ranged in the coordinate space.
Specifically, the limiting fully-correlated case of the interacting version of Burin-Maksimov model was recently analyzed in \cite{Yao_Lukin2014,Moessner2017,luitz2018emergent,Botzung_Muller2018,Botzung_Muller2018,Roy2019,Nag2019}, and the opposite situation without any constraints was considered in details in \cite{Burin_Kagan1994,BurinPRB2015-1,BurinPRB2015-2,BurinAnnPhys2017,tikhonov2018many,deTomasi2018}. However, the intermediate regime represented by both finite means and dispersion in distribution functions of matrix elements
needs deep consideration.

Finally, in order to more precisely investigate the role of partial correlations in long-range random matrix models, it would be insightful to construct the corresponding effective field-theoretic description with imposed soft constraints. For instance, this approach may
naturally incorporate ergodicity-breaking phenomena in gauge-invariant lattice models~\cite{Heyl2018gauge_fields}. Although the desirable theory is more technically involved than usual super-symmetric non-linear sigma model due to finite means of hopping and effective non-locality, several attempts were made to describe systems with correlations in off-diagonal terms~\cite{Ossipov2013, Ossipov2006}. Moreover, so-called ``virial expansion'' for almost diagonal random matrices developed in~\cite{Yevtushenko_2003, Yevtushenko_2007} seems to be a suitable candidate for an appropriate representation of constrained models with disjoint sub-blocks in Hilbert space.

\begin{acknowledgments}
We are grateful to 
V.~E.~Kravtsov for stimulating discussions.
P. N. appreciates warm hospitality of the Max-Planck Institute for the Physics of Complex Systems, Dresden, Germany, extended to him during his visits when this work was done.
P.~N. acknowledges funding by the RFBR, Grant No. 17-52-50013, and the Foundation for the Advancement to Theoretical Physics and Mathematics BASIS Grant No. 17-11-107.
I.~M.~K. acknowledges the support of German Research Foundation (DFG) Grant No. KH~425/1-1 and the Russian Foundation for Basic Research Grant No. 17-52-12044.
\end{acknowledgments}

\bibliography{Lib}

\begin{thebibliography}{94}
\bibitem [{On-()}]{On-site-disorder-footnote}%
  \BibitemOpen
  \href@noop {} {}\bibinfo {note} {Here and further we consider the systems
  only with the uncorrelated diagonal disorder and focus on independent
  Gaussian distributed on-site energies $\epsilon_n$ with zero mean $<
  \epsilon_n>$ and the variance $< \epsilon_n>^2=W^2$.}\BibitemShut {Stop}%
\end{thebibliography}

\appendix

\section{Yuzbashyan-Shastry model}\label{App_Sec:YS}
\subsection{Model}
Consider an ensemble of random $N\times N$ Hermitian matrices
\begin{gather}\label{App:ham}
H_0 =\diag \ep+j^0 g^T g^* \ ,
\end{gather}
with a random real vector $\ep = (\ep_1,\ldots,\ep_N)$ and deterministic complex vector $g = (g_1,\ldots,g_N)$ with $N$-independent elements.
$\ep_i$ are statistically independent entries with a zero mean and a unit variance
\begin{gather}\label{App:ep_distr}
\langle \ep_{k}\rangle = 0, \quad \langle \ep_{k}^2\rangle = 1,
\end{gather}
while $j^0$ scales with the matrix size $N$ as $j^0 \propto N^{-\gamma_0/2}$.
Note that the translation-invariant case of $g = (1,\ldots,1)$ corresponds to the finite mean $j^0$ of the hopping elements (plus additional unimportant shift of energy).

As Type-1 Hamiltonian~\cite{Yuzbashyan_JPhysA2009_Exact_solution,Yuzbashyan_NJP2016} $H_0$ provides an exact eigenproblem solution \cite{Yuzbashyan_JPhysA2009_Exact_solution}
\begin{gather}\label{App:H0_eigenstates}
\psi_{E_n}(k) = C_n \frac{g_k^*}{E_n - \ep_k} \ ,
\end{gather}
with the (possibly $N$-dependent) normalization constant $C_n = j^0\sum_m g_m \psi_{E_n}(m)$,
\be\label{App:norm_cond}
|C_n|^{-2} = \sum_k \frac{|g_k|^2}{(E_n - \ep_k)^2} \ ,
\ee
 and the secular equation for the spectrum $E=E_n$
\begin{gather}\label{App:Spectral_eq_E}
\sum_m \frac{|g_m|^2}{E-\ep_m} = \frac{1}{j^0}\sim N^{\gamma_0/2} \ ,
\end{gather}
giving all eigenvalues (except the highest one $E_N>\ep_N$) lying between adjacent bare levels
\begin{gather}\label{App:ep_m<E_m<ep_m+1}
\ep_m<E_m<\ep_{m+1} \ .
\end{gather}
Here we assumed $\ep_m$ to be ordered in ascending order $\ep_m<\ep_{m+1}$, $m=\overline{1,N-1}$.
Note that this model includes the limiting case $a=0$ of long-range deterministic hoppings considered in \cite{Kravtsov_Shlyapnikov_PRL2018}
both for positive ($g_m = 1$) and staggering ($g_m = (-1)^m$) hopping elements.

\subsection{Spectral statistics}
To find the spectrum $E_n$ one should consider the secular equation in more details.
Let's assume that all $g_k$ are of the same order $g_k\simeq g\sim N^0$ and consider the variation of $E_n$ from its bare value
$\ep_n$ as
\begin{gather}\label{App:dE_m<ep_m+1-ep_m}
\Gamma_n = E_n-\ep_n, \quad 0\leq \Gamma_n< \ep_{n+1}-\ep_{n} \ .
\end{gather}

Separating positive and negative summands of the sum \eqref{App:Spectral_eq_E} for $E=E_n$
we obtain
\begin{multline}\label{App:sum_eqp_m}
\frac{1}{j^0} = \frac{|g_n|^2}{\Gamma_n}+\sum_{k>0} \left[\frac{|g_{n-k}|^2}{\ep_n -\ep_{n-k}+\Gamma_n}-\frac{|g_{n+k}|^2}{\ep_{n+k} -\ep_n-\Gamma_n}\right]
\\ \simeq
\frac{|g|^2}{\Gamma_n}+\sum_{k>0} \left[\frac{|g|^2}{\ep_n -\ep_{n-k}+\Gamma_n}-\frac{|g|^2}{\ep_{n+k} -\ep_n-\Gamma_n}\right] \ .
\end{multline}

Now we have to estimate a typical value of the sum taking into account the inequality \eqref{App:dE_m<ep_m+1-ep_m}.
To do so we consider two limiting cases.

(i)~In the limit of a completely rigid spectrum
$\ep_n = n\delta$,
with the mean level spacing $\delta \sim 1/[\rho(0) N]$ and the density of states at the Fermi level $\rho(0)$,
the sum \eqref{App:sum_eqp_m} can be taken explicitly
\be
\frac{1}{|g|^2 j^0} = \frac{\pi}{\delta \tan(\pi \Gamma_n/\delta)}
\ee
giving
\be\label{App:dE_n_res}
\Gamma_n = \frac{\delta}{\pi}\arctan(\pi |g|^2 j^0/\delta) \sim N^{-\gamma_{\rm YS}/2} \ ,
\ee
where
\be\label{App:gamma_eff}
\gamma_{\rm YS} = \max(\gamma_0,2) \ .
\ee

Analogously the normalization constant governed by \eqref{App:norm_cond} takes the form
\begin{multline}\label{App:norm_coef}
\frac{1}{|g|^2|C_n|^2} \simeq \sum_k \frac{1}{(\Gamma_n +\ep_n- \ep_k)^2} \\
\simeq
\frac{1}{\Gamma_n^2}+\frac{\pi^2}{\delta^2\sin^2(\pi \Gamma_n/\delta)}\simeq \frac{1}{\Gamma_n^2}
\ .
\end{multline}
In the latter equality we neglect prefactors, focus only on the $N$-scaling and take into account that $\Gamma_n\lesssim \delta$.

(ii)~In the opposite limit of uncorrelated eigenstates one can calculate \eqref{App:sum_eqp_m} as follows
\begin{multline}\label{App:Gamma_n_uncorr}
\frac{1}{|g|^2 j^0} =
\frac{1}{\Gamma_n}+N \int_{|\omega|>\delta} \frac{\rho(\ep_n-\omega)d\omega}{\omega+\Gamma_n}\simeq\\
\frac{1}{\Gamma_n}+
N \rho(0)\int_{\delta}^{\infty} d\omega\left[\frac{1}{\omega+\Gamma_n}-\frac{1}{\omega-\Gamma_n}\right] =\\
\frac{1}{\Gamma_n}+\frac{1}{\delta} \ln\left(\frac{\delta-\Gamma_n}{\delta+\Gamma_n}\right)
 \ ,
\end{multline}
where we take into account \eqref{App:dE_m<ep_m+1-ep_m} in the lower limits of integration.
Here the density of states is
\be
\rho(\ep) = \frac{1}{N}\sum_k \la \delta(\ep-\ep_k)\ra \ .
\ee

Again considering only $N$-scaling of $\Gamma_n$ in \eqref{App:Gamma_n_uncorr} one can obtain
\begin{gather}\label{App:dE_n_res_uncorr}
\Gamma_n \sim N^{-\gamma_{\rm YS}/2} \ .
\end{gather}

Analogously the normalization constant governed by \eqref{App:norm_cond} takes the form of \eqref{App:norm_coef}
\begin{multline}\label{App:norm_coef_uncorr}
\frac{1}{|g|^2|C_n|^2}
\frac{1}{\Gamma_n^2}+N \rho(0)\int_{\delta}^{\infty} d\omega\left[\frac{1}{(\omega+\Gamma_n)^2}-\frac{1}{(\omega-\Gamma_n)^2}\right] =\\
\frac{1}{\Gamma_n^2}+\frac{1}{\delta} \left(\frac1{\delta+\Gamma_n}-\frac1{\delta-\Gamma_n}\right)\simeq \frac{1}{\Gamma_n^2}
\ .
\end{multline}

As the scaling~(\ref{App:dE_n_res},~\ref{App:dE_n_res_uncorr}) of the energy deviation $\Gamma_n$~\eqref{App:dE_m<ep_m+1-ep_m}
and of the normalization constant $C_n$~(\ref{App:norm_coef},~\ref{App:norm_coef_uncorr}) are the same in both limiting cases, we conclude that these parameters weakly depends on the statistics of bare levels $\ep_n$.

\subsection{Eigenstate statistics}
Using the results \eqref{App:H0_eigenstates} and \eqref{App:norm_coef} one can calculate the spectrum of fractal dimensions $f(\alpha)$ for the wave function intensity
\be
|\psi_{E_n}(k)|^2 =  \frac{|C_n|^2 |g|^2}{(E_n - \ep_k)^2}\simeq \frac{1}{[(\ep_n-\ep_k)N^{\gamma_{\rm YS}/2}-1]^2}\equiv N^{-\alpha} \ .
\ee

Indeed, as the probability of $\ep_n - \ep_k \sim N^{-p/2}$ for $p>0$ is
\be
dP(\ep_n - \ep_k \sim N^{-p/2})\sim P(p)dp \sim N^{-p/2}dp \ ,
\ee
one can easily find
\be
\alpha=\left\{
\phantom{0}\gamma_{\rm YS}-p, \quad p<\gamma_{\rm YS}\atop
\phantom{\gamma_{\rm YS}-p}0, \quad p>\gamma_{\rm YS}
\right.
\ee
and
\be
N^{f(\alpha)-1} = dP(p(\alpha))\sim N^{-p(\alpha)/2}=N^{(\alpha-\gamma_{\rm YS})/2}
\ee
for $p=\gamma_{\rm YS}-\alpha>0$,
giving
the spectrum of fractal dimensions of the form of \cite{Kravtsov_NJP2015}
\be\label{App:f(a)_res}
f(\alpha) = 1+\frac{\alpha-\gamma_{\rm YS}}{2} \ , \quad 0<\alpha<\gamma_{\rm YS} \ ,
\ee
with $\gamma_{\rm YS}$ given by \eqref{App:gamma_eff}.

As a result, unlike the Rosenzweig-Porter model,
the YS model \eqref{App:ham} shows only localized
and critical wave functions of {\it all} eigenstates, except the only top energy state at $\gamma_0<2$~\cite{Yuzbashyan_NJP2016}.
At $\gamma_0>2$ the wave-function statistics~\eqref{App:f(a)_res} coincides with the one of the RP, with $\gamma_{\rm YS}=\gamma_0$, while
at all $\gamma_0<2$ instead of the delocalized phases YS model shows the critical localization with $\gamma_{\rm YS}=2$.

\section{Matrix inversion trick}\label{App_Sec:Matrix_inversion}
In this Appendix we first give general estimate of the optimal parameter $\beta$, $E_0\sim N^{\beta}$ and then
apply the matrix inversion trick~\cite{Nosov2019correlation} to the PLRBM model with partial correlations~\eqref{eq:part_corr}

\subsection{Optimization over $E_0$ in the matrix inversion trick}
Let's focus for simplicity on the case of $E \sim W\sim N^0$.
The effective hopping matrix $J_{m-n} = M_{m-n}(E+E_0-\ep_n)/M_0$ in Eq.~\eqref{eq:Matrix_inversion_0}
\be
\left[(1+\hat j^0/E_0)^{-1}(E-\hat\ep+E_0)-E_0\right]\left|\psi_{E}\ra = 0 \ .
\ee
can be estimated as follows
\be\label{App:J_in_p}
\hat J = \sum_p \lv p\ra \la p \rv \frac{E+E_0}{M_0 (E_0+j^0_{p})}\equiv \sum_p J_p\lv p\ra \la p \rv
\ee
Here we neglected the term $\ep_n$ for simplicity as $E \sim \ep_n \sim N^0$
and divided Eq.~\eqref{eq:Matrix_inversion_0} by
\be
M_0 = \frac{1}{N}\sum_p \frac{1}{E_0+j^0_{p}} \simeq \frac{1}{E_0 + j^0_{p,\min}}
\ee
in order to have diagonal disorder in the standard form of $\ep_n$.
Here $j_{p,\min}^0$ is the typical hopping energy level and $E_0$ is taken to be in the gap of the hopping spectrum or beyond it (see Fig.~\ref{Fig:j_p_spectrum}).

Each term in the sum~\eqref{App:J_in_p} can be minimized over $E_0$ giving
\be
E_0 \sim \min (|j^0_{p,\min}|, |E|), \quad J_p \simeq \frac{E |j^0_{p,\min}|}{j^0_{p,\min}+j^0_{p}} \ .
\ee
However, one should take into account that $E_0$ has to be beyond the spectrum $|E_0|\gtrsim |j^0_{p,\min}|$, leading to
the final result
\be
E_0 \simeq j^0_{p,\min}, \quad J_p \simeq \frac{(E+j^0_{p,\min})j^0_{p,\min}}{j^0_{p,\min}+j^0_{p}} \ .
\ee
In the case of BM or YS deterministic hopping $\hat j^0$, the typical energy level $j^0_{p,\min} \sim N^0$, thus,
the optimal $E_0\sim N^0$ and this confirms the statement given in the main text.

\subsection{Matrix inversion for PLRBM model with partial correlations}
We start with the model~(\ref{eq:ham},~\ref{eq:j_def})
with $\tilde j_{mn} = |m-n|^{-\tilde a}$, $j^0_{mn} = |m-n|^{-a_0}$
and compute matrix elements $\tilde{j}^0_{mn} + r_{mn}$
of the effective Hamiltonian \eqref{eq:Matrix_inversion_result}
\be
\hat{\tilde j}^0 +\hat{\tilde j} + \hat r=\left(1+\hat{j}^0/E_0\right)^{-1} \left(E-\hat \ep +E_0+\hat{\tilde j}\right) \ .
\ee
The first terms in the brackets of r.h.s. $\left(1+\hat{j}^0/E_0\right)^{-1} \left(E-\hat \ep +E_0\right)$ scales as $|n|^{-(2-a)}$ at $a<1$ and correspond to $(1-\alpha)\hat{\tilde j}^0$ with a certain constant $\alpha$ of order of one.
The calculation of it is given in~\cite{Nosov2019correlation}.
Further we consider the rest part
\be
\alpha\hat{\tilde j}^0 +\hat{\tilde j} + \hat r\equiv \left(1+\hat{j}^0/E_0\right)^{-1} \hat{\tilde j} \ .
\ee

In order to simplify the calculations we provide the upper bound of this term
by replacing the oscillating amplitudes $h_{mn}$ of $\tilde j_{mn} = h_{mn}/|m-n|^{\tilde a}$ by their maximal absolute value $h_{mn}=1$.
Within this approximation the result can be easily derived in the momentum space, since both matrices are diagonal in that basis. For purposes of clarity we reproduce main steps of similar calculations presented in \cite{Nosov2019correlation} specifically for the case of our interest.

We start by writing down the Fourier-transformed hopping amplitude $j^0$ (in case of $\tilde j$ one needs to replace $a_0$ by $\tilde a$) in different asymptotic regimes
\be
 j^0_p/2 \simeq \zeta_{a_0} + A_{a_0}
\left(\frac{N}{|p|}\right)^{1-a_{0}},\; \text{ for } |p|\ll N \ ,
\ee
\be
 j^0_p/2 \simeq j^0_{\min} + B_{a_0}
\left(\frac{2q}{N}\right)^{2}, \;\text{ for } q=|N/2-p|\ll N \ ,
\ee
where the corresponding constants are given for $a_0>0$ by
\be
A_{a_0} = (2\pi)^{a_0-1}\Gamma_{1-a_0} \sin \frac{\pi a_0}{2} \ ,
\ee
for $a_0\ne 2m+1, m \in \mathbb{N}$ and
\begin{align}
j^0_{\min} &= 2  (2^{1-a_0}-1)\zeta_{a_0}<0\ , \\
B_{a_0} &= 8 \pi^2 (1-2^{3-a_0})\zeta_{a_0-2}\simeq 2\pi^2  a_0>0\ .
\end{align}
Here  $\zeta_{a_0}$ is the Riemann zeta function.

The next step is to estimate the long-range asymptotic behavior of the effective Hamiltonian $(1-\alpha)\hat{\tilde j}^0+\hat{\tilde j} + \hat r$ given by
\be \label{App:effective_j}
(1-\alpha){\tilde j}^0_n+{\tilde j}_n + r_n=
C_0(a_0,\tilde a)+\frac{2}{N}\text{Re}\sum\limits_{p=1}^{N/2}\frac{\tilde{j}_p\;e^{2\pi i p n /N}}{E_0+j^0_p}
\ee
with the zero-momentum contribution
\be
C_0(a_0,\tilde a)=\frac{\zeta_{\tilde a}+N^{1-\tilde a}/(1-\tilde a)}{N(E_0/2 +\zeta_{a_0})+N^{2-a_0}/(1-a_0)} \ .
\ee
 The last term in \eqref{App:effective_j} can be split into three parts corresponding to different regimes of $\tilde j_{p}$ and $ j^0_{p}$. However, not all of the resulting terms are equally important in the limit $1\ll n \ll N$. One can easily show that summation only over sufficiently small momenta $|p|<\alpha N$ (where $0<\alpha<1/2$) contributes to the long-range behavior of matrix elements. The rest of the summation results in the gives small contribution to the residue term $r_n$ and is unimportant. Thus, we focus only on the following sum
\begin{multline}
S(a_0,\tilde a)= \frac{2 \zeta_{\tilde a}}{N}\text{Re}\sum\limits_{p=1}^{N\alpha_1}\frac{e^{2\pi i p n /N}}{E_0/2 +\zeta_{a_0} +A_{a_0} (p/N)^{a_0-1} }\\+
\frac{2 A_{\tilde a}}{N}\text{Re}\sum\limits_{p=1}^{N\alpha_1}\frac{(p/N)^{\tilde a-1}e^{2\pi i p n /N} }{E_0/2 +\zeta_{a_0} +A_{a_0} (p/N)^{a_0-1} }\ .
\end{multline}


In the case of our main interest ($a_0<1$) an additional momentum scale $p_c=N[E_0/2 A_{a_0}+\zeta_{a_0}/A_{a_0}]^{-1/(1-a_0)}\leq N$ emerges, and for $p<p_c$ the denominator is represented by the power-law contribution. Contrary, for $p>p_c$ a constant term is dominant and we get
\begin{multline}\label{App:S(N)}
S(N\alpha,a_0<1,\tilde a)\approx \frac{ \zeta_{\tilde a} A_{a_0-1}}{\pi A_{a_0}}\frac{1}{|n|^{2-a_0}}\\
+\frac{ A_{\tilde a} A_{1-\tilde a}}{\pi(E_0/2+\zeta_{a_0})} \;\frac{1}{|n|^{\tilde a}}
+\frac{ A_{\tilde a}A_{a_0-\tilde a}}{\pi  A_{a_0}}\frac{1}{|n|^{1-a_0+\tilde a}}\ ,
\end{multline}
which for $a_0<1$ leads to the effective parameter governing the long-range tails of typical wave functions:
\be
a_{\rm eff} = \min\left(\tilde a, 2-a_0\right)
\ee
in full agreement with the result \eqref{eq:a_eff_PLRBM+BM} mentioned in the main text.
The last term in~\eqref{App:S(N)} contribute to the residual term as it is small compared to the second term $\sim|n|^{-\tilde a}$ in the considered interval $a_0<1$.

\section{Self-consistent method}\label{App_Sec:self-consist}
In this Appendix we use the formulation of the eigenproblem in terms of exact Eqs.~(\ref{eq:psi_n(m)_dE_n_solution}~-~\ref{eq:a_Ps_self-consist})
in order to derive expressions for
scaling~\eqref{eq:Gamma_n_res} of the broadening factor $\Gamma_n$ in the average wave-function intensity~\eqref{eq:psi2_in_P_mn}.

In the first part we restrict our consideration to the first and second moments of the parameters $a_n$, $P_{kn}$,
calculate sums $S_1$~\eqref{eq:S_1_res} and $S_2$ (see below) in two limiting cases of
(i)~the completely rigid spectrum $\ep_n -\ep_m = (n-m)\delta$, and
(ii)~the Poisson statistics of $\ep_n$,
derive the mean expressions~\eqref{eq:a_Ps_mean} for the RP model with partial correlations and the effective expression~\eqref{eq:Gamma_n_res} for the broadening parameter $\Gamma_n$ depending solely on $S_1$.
Next, we explicitly calculate $N$-scaling~\eqref{eq:Gamma_n_res} of $\Gamma_n$.


\subsection{Decoupling of correlations}

In order to solve Eqs.~\eqref{eq:a_Ps_self-consist} we assume that
$P_{kn}$ and $a_n$ are uncorrelated from each other and from $h_{lm}$
and calculate the mean and the variance for each of them averaging over off-diagonal matrix elements $h_{kn}=h_{nk}$ assumed to be real and taking into account $\la h_{lm}\ra = 0$, $\la h_{lm}^2\ra = 1$.
As a result for $k\ne n$,
\ben
\la P_{kn} \ra = \sum_{l\ne n} \tilde j \la h_{kl}\ra \frac{\la a_n + P_{ln} + \tilde j  h_{ln}\ra }{\omega_{nl} + \Gamma_n} = 0\ ,
\een
\ben
\la P_{nn} \ra = \sum_{l\ne n} \frac{\tilde j \la h_{nl} \ra \la a_n + P_{ln}\ra  + \tilde j^2  \la h_{ln}^2\ra }{\omega_{nl} + \Gamma_n}
\equiv \tilde j^2  S_1 \ ,
\een
\ben
\frac{\la a_n \ra}{j^0}  = 1+\sum_{l\ne n} \frac{\la a_n \ra + \tilde j \la h_{ln}\ra + \la P_{ln}\ra}{\omega_{nl} + \Gamma_n} \equiv 1 + \la a_n \ra S_1 \ ,
\een
\begin{multline*}
\la P_{kn}^2 \ra = \\
\tilde j^2 \sum_{l,l'\ne n} \la h_{kl} h_{kl'}\ra \frac{\la (a_n + P_{ln} + \tilde j  h_{ln})(a_n + P_{l'n} + \tilde j  h_{l'n})\ra }{(\omega_{nl} + \Gamma_n)(\omega_{nl'} + \Gamma_n)}
\\ = \tilde j^2 \sum_{l\ne n} \frac{\la a_n^2 \ra + \la P_{ln}^2\ra + \tilde j^2  \la h_{ln}^2\ra}{(\omega_{nl} + \Gamma_n)^2}
\\ \equiv \tilde j^2  S_2\lp\la a_n \ra^2+\sigma_a^2 + \la P_{ln}^2\ra + \tilde j^2 \rp \ ,
\end{multline*}
\begin{multline*}
\sigma_P^2 = \la P_{nn}^2\ra - \la P_{nn}\ra^2 = \\
\tilde j^2 \sum_{l,l'\ne n} \la h_{nl} h_{nl'}\frac{(a_n + P_{ln} + \tilde j  h_{ln})(a_n + P_{l'n} + \tilde j  h_{l'n}) }{(\omega_{nl} + \Gamma_n)(\omega_{nl'} + \Gamma_n)}\ra -\tilde j^4 S_1^2\\=
\tilde j^2 \sum_{l\ne n} \frac{\la a_n^2\ra + \la P_{ln}^2\ra + 3\tilde j^2 }{(\omega_{nl} + \Gamma_n)^2} +
\sum_{l\ne l'\ne n} \frac{\tilde j^4}{(\omega_{nl} + \Gamma_n)(\omega_{nl'} + \Gamma_n)} -\tilde j^4 S_1^2 \\ =
\tilde j^2  S_2\lp\la a_n\ra^2+\sigma_a^2 + \la P_{ln}^2\ra + 2\tilde j^2 \rp
 \ ,
\end{multline*}
\begin{multline*}
\frac{\sigma_a} {(j^0)^{2}}  = \frac{\la a_{n}^2\ra - \la a_{n}\ra^2}{(j^0)^{2}}  =\\
\sum_{l,l'\ne n} \la \frac{(a_n + P_{ln} + \tilde j  h_{ln})(a_n + P_{l'n} + \tilde j  h_{l'n}) }{(\omega_{nl} + \Gamma_n)(\omega_{nl'} + \Gamma_n)}\ra -\la a_n-j^0\ra^2=\\
\sum_{l\ne n} \frac{\la P_{ln}^2\ra + \tilde j^2 }{(\omega_{nl} + \Gamma_n)^2} +
\sum_{l,l'\ne n} \frac{\la a_n^2\ra-\la a_n\ra^2}{(\omega_{nl} + \Gamma_n)(\omega_{nl'} + \Gamma_n)} =\\
\lp\la P_{ln}^2\ra + \tilde j^2 \rp S_2 +\sigma_a^2 S_1^2
\ .
\end{multline*}
Here and further we use the notation $\omega_{nl} = \ep_n-\ep_l$ for brevity.

This gives the following self-consistency equations
\bes\label{App:mean+var_a_Ps}
\begin{align}
\la P_{kn} \ra &= \tilde j^2  S_1 \delta_{kn} \ , \\
\la a_n\ra &= \frac{1}{(j^0)^{-1}-S_1} \sim \min (j_0, S_1^{-1})\ , \\
\la P_{kn}^2 \ra &= \frac{\la a_n \ra^2+\sigma_a^2 + \tilde j^2 }{1-(\tilde j^2  S_2)^{-1}}\nonumber \\
&\sim\lp\la a_n \ra^2+\sigma_a^2 + \tilde j^2\rp \min\lp1,\tilde j^2  S_2\rp
\ ,\\
\sigma_P^2 &=  \la P_{kn}^2 \ra+\tilde j^4 S_2 \ , \\
\sigma_a^2  &= \frac{\lp\la P_{ln}^2\ra + \tilde j^2 \rp S_2}{(j^0)^{-2}-S_1^2}\nonumber \\
&\sim \frac{S_2}{S_1^2}\lp\la P_{ln}^2\ra + \tilde j^2 \rp \min\lp 1,(j^0 S_1)^{2}\rp\ ,
\end{align}
\ees
with the sums
\ben\label{App:S_1,2}
S_1 = \sum_{l\ne n} \frac{1}{\omega_{nl} + \Gamma_n}, \;
S_2 = \sum_{l\ne n} \frac{1}{(\omega_{nl} + \Gamma_n)^2} \ .
\een

While averaging the denominators $\omega_{nl}+\Gamma_n$ we estimate only $N$-scaling of the broadening parameter as follows
\be\label{App:Gamma_n}
\Gamma_n \sim \tilde j +\la a_n\ra +\la P_{nn}\ra + \sigma_a + \sigma_P \
\ee
and we consider the typical energy position $\ep_n$ to lie a bit asymmetrically in the middle of the spectrum, thus the summation
would be in the limits $W_1<\omega_{nl}<W_2$, with $|W_1-W_2| \sim O(1)$ and $W_1 + W_2 = W$.

To estimate a typical value of the sums \eqref{eq:S_1,2} we consider two limiting cases.

(i)~In the limit of the completely rigid spectrum $\ep_n = n\delta$, with the mean bare level spacing $\delta \sim 1/[\rho(0) N]\sim W/N$ and the density of states at the Fermi level $\rho(0)$,
the sums can be taken explicitly
\ben
S_1 = \left\{
\begin{array}{ll}
\frac{\ln(W_2/W_1)}{\delta}+\frac{\pi}{\delta\tan(\pi \Gamma_n/\delta)}-\frac{1}{\Gamma_n},
& \Gamma_n\ll W\\
\frac{N}{\Gamma_n}, &  \Gamma_n\gg W\\
\end{array}
\right.
\een
\ben
S_2 = \left\{
\begin{array}{ll}
\frac{\pi^2}{\delta^2}+\frac{\pi^2}{\delta^2 \sin^2(\pi \Gamma_n /\delta)}-\frac{1}{\Gamma_n^2}, & \Gamma_n\ll W\\
\frac{N}{\Gamma_n^2}, &  \Gamma_n\gg W \\
\end{array}
\right. \
\een
and provide the following asymptotics
\bes\label{App:S_1,2_res_rigid}
\begin{align}
S_1 &\sim \left\{
\begin{array}{ll}
\frac{1}{\delta},
& \Gamma_n\ll W\\
\frac{N}{\Gamma_n}, &  \Gamma_n\gg W\\
\end{array}
\right.
\\
S_2 &\sim \left\{
\begin{array}{ll}
\frac{1}{\delta^2}, & \Gamma_n\ll W\\
\frac{N}{\Gamma_n^2}, &  \Gamma_n\gg W \\
\end{array}
\right. \
\end{align}
\ees

(ii)~In the opposite limit of uncorrelated eigenstates one can calculate \eqref{eq:S_1,2} as follows

\begin{multline*}
S_1 = \\
\frac{N}{2} \lb\int_{-W_1}^{-\delta} \frac{\rho(\ep_n-\omega)d\omega}{\omega+\Gamma_n}+\int_{\delta}^{W_2} \frac{\rho(\ep_n-\omega)d\omega}{\omega+\Gamma_n}\rb\simeq\\
\frac{N \rho(0)}{2}\left[\int_{\delta}^{W_1} \frac{d\omega}{\omega+\Gamma_n}-\int_{\delta}^{W_2} \frac{d\omega}{\omega-\Gamma_n}\right] =\\
\frac{1}{2\delta} \lb\ln\lv\frac{W_1+\Gamma_n}{W_2-\Gamma_n}\rv-\ln\lv\frac{\delta-\Gamma_n}{\delta+\Gamma_n}\rv\rb
 \ .
\end{multline*}
\begin{multline*}
S_2 = \\
\frac{N}{2} \lb\int_{-W_1}^{-\delta} \frac{\rho(\ep_n-\omega)d\omega}{(\omega+\Gamma_n)^2}+\int_{\delta}^{W_2} \frac{\rho(\ep_n-\omega)d\omega}{(\omega+\Gamma_n)^2}\rb\simeq\\
\frac{N \rho(0)}{2}\left[\int_{\delta}^{W_1} \frac{d\omega}{(\omega+\Gamma_n)^2}+\int_{\delta}^{W_2} \frac{d\omega}{(\omega-\Gamma_n)^2}\right] =\\
\frac{N}{(\Gamma_n-W_1)(\Gamma_n-W_2)}-\frac{1}{\Gamma_n^2-\delta^2}
 \ .
\end{multline*}
Unlike YS model~\eqref{App:Gamma_n_uncorr} here the broadening parameter can be both smaller and larger than bare mean level spacing $\delta$, thus,
during the calculation we just take into account the fact that $E_n - \ep_k = \omega +\Gamma_n$ is off-resonant. 
Here the density of states is
$
\rho(\ep) = \sum_k \la \delta(\ep-\ep_k)\ra/N \ .
$

In this case asymptotics read as follows
\bes\label{App:S_1,2_res_Poisson}
\begin{align}
S_1 &\sim \left\{
\begin{array}{ll}
\frac{1}{\delta}, & \Gamma_n\ll W\\
\frac{N}{\Gamma_n}, &  \Gamma_n\gg W\\
\end{array}
\right.
\\
S_2 &\sim \left\{
\begin{array}{ll}
\frac{1}{\delta^2},
& \Gamma_n\ll \delta\\
\frac{1}{\Gamma_n^2}+\frac{1}{\delta},
& \delta\ll\Gamma_n\ll W\\
\frac{N}{\Gamma_n^2}, &  \Gamma_n\gg W \\
\end{array}
\right. \
\end{align}
\ees

Note that the expressions for sum $S_1$ are the same in both cases, while $S_2$ are different only in the non-ergodic extended phase.
Let's show that this difference do not affect the result~\eqref{eq:Gamma_n_res} for the broadening parameter $\Gamma_n$.
In order to prove it we consider the expression~\eqref{App:Gamma_n} in more details substituting the expressions~\eqref{App:mean+var_a_Ps}
one by one.

First, let's substitute $\sigma_P$
\begin{multline*}
\Gamma_n \sim \tilde j +\la a_n\ra +\la P_{nn}\ra + \sigma_a + \tilde j^2 \sqrt{S_2}+\\
\lp\la a_n \ra+\sigma_a + \tilde j \rp\min\lp 1, \tilde j  \sqrt{S_2}\rp \ .
\end{multline*}
As $\min\lp 1, \tilde j  \sqrt{S_2}\rp \leq 1$ one can neglect the whole last summand corresponding to $\sqrt{\la P_{kn}^2\ra}$ which not larger than $\la a_n \ra+\sigma_a + \tilde j$.
Next, we substitute $\sigma_a$
\begin{multline*}
\Gamma_n \sim \tilde j +\la a_n\ra +\la P_{nn}\ra + \\
\frac{\sqrt{S_2}}{S_1}\lp\sqrt{\la P_{kn}^2\ra} + \tilde j \rp \min\lp 1,j^0 S_1\rp + \tilde j^2 \sqrt{S_2} \ .
\end{multline*}

As $S_1^2\geq S_2$ in all phases of both limiting cases,
one can neglect
$\tilde j^2 \sqrt{S_2}$ comparing to $\la P_{nn}\ra \sim \tilde j^2 S_1$
and the whole $\sigma_a$ term comparing to $\la a_n \ra+\tilde j$
as $\lp\sqrt{S_2}/{S_1}\rp\min\lp 1,j^0 S_1\rp \leq 1$.
As a result we come to the expression~\eqref{eq:Gamma_n_res}
\be\label{App:Gamma_n_res}
\Gamma_n \sim \tilde j +\la a_n\ra +\la P_{nn}\ra \sim \tilde j + \min (j_0, S_1^{-1}) + \tilde j^2 S_1 \ ,
\ee
depending solely on $S_1$.
This confirms the statement given in the main text and concludes this section.


\subsection{Calculation of $N$-scaling~\eqref{eq:Gamma_n_res} of $\Gamma_n$}
Using Eq.~\eqref{App:Gamma_n_res} and expressions~(\ref{App:S_1,2_res_rigid}, \ref{App:S_1,2_res_Poisson}) for $S_1$,
in this section we calculate the scaling \eqref{eq:Gamma_n_res} of $\Gamma_n$ in all three phases:

\begin{enumerate}
  \item $\Gamma_n\gg W$. In this case $S_1 = N/\Gamma_n$ and Eq.~\eqref{App:Gamma_n_res} takes the form
\be
\Gamma_n \sim N^{-\tilde \gamma/2} + \min \lp N^{-\gamma_0/2}, \frac{\Gamma_n}N\rp + \frac{N^{1-\tilde \gamma}}{\Gamma_n} \ ,
\ee
The second term does not play any role as it is less than $\Gamma_n/N\ll\Gamma_n$, while the third term dominates over the first one and gives
$\Gamma_n\sim N^{(1-\tilde \gamma)/2}$ and thus $\tilde\gamma < 1$.

  \item $\delta\ll\Gamma_n\ll W$. In this case $S_1 = 1/\delta \sim N$ leading to
\be
\Gamma_n \sim N^{-\tilde \gamma/2} + \min \lp N^{-\gamma_0/2}, \delta\rp + N^{1-\tilde \gamma} \ .
\ee
As in the previous case the second term does not play any role as it is less than $\delta\ll \Gamma_n$, while the third term dominates over the first one and gives
$\Gamma_n\sim N^{1-\tilde \gamma}$ and thus $1<\tilde\gamma < 2$.


  \item $\Gamma_n\ll \delta$. Here $S_1 \sim 1/\delta$ leading to $\tilde \gamma>2$ and
\be
\Gamma_n \sim N^{-\tilde \gamma/2} + \min \lp N^{-\gamma_0/2}, \delta\rp + N^{1-\tilde \gamma} \ .
\ee
In this case the first term dominates over the third one and the concurrence of the first two terms gives the desired result
$\Gamma_n\sim N^{-\tilde \gamma/2}+N^{-\gamma_{\rm YS}/2}\sim N^{-\gamma_{\rm eff}/2}$, with $\gamma_{\rm eff}$ given by~\eqref{eq:gamma_eff_RP+YS}.

\end{enumerate}


\end{document}